\documentclass{article}

\usepackage{arxiv}

\usepackage[utf8]{inputenc} % allow utf-8 input
\usepackage[T1]{fontenc} % use 8-bit T1 fonts
\usepackage[colorlinks=true, allcolors=blue]{hyperref}     % hyperlinks
\usepackage{url}            % simple URL typesetting
\usepackage{booktabs}       % professional-quality tables
\usepackage{amsfonts}       % blackboard math symbols
\usepackage{amsmath}
\usepackage{nicefrac}       % compact symbols for 1/2, etc.
\usepackage{microtype}      % microtypography
\usepackage{lipsum}
\usepackage{graphicx}
\usepackage{xcolor}
\usepackage{enumitem}
\usepackage[nottoc,numbib]{tocbibind}
\usepackage{floatrow}

\usepackage{tikz}
\usetikzlibrary{arrows,arrows.meta,positioning,automata} 
\tikzset{
    >=stealth',
    punkt/.style={
           rectangle,
           rounded corners,
           draw=black, very thick,
           text width=6.5em,
           minimum height=2em,
           text centered},
    pil/.style={
           ->,
           thick,
           shorten <=2pt,
           shorten >=2pt,}
}
\usepackage{circuitikz}

\definecolor{darkgray}{rgb}{0.5,0.5,0.5}
\definecolor{lightgray}{rgb}{0.95,0.95,0.95}

\newcommand\hide[1]{}
\newboolean{show_todos}
\newboolean{show_notes}
\newboolean{show_feedback}
\newboolean{show_contributions}

\setboolean{show_todos}{true}
\setboolean{show_notes}{true}
\setboolean{show_feedback}{true}
\setboolean{show_contributions}{false}

\newcommand\todo[1]{\ifthenelse{\boolean{show_todos}}{\textcolor{red}{\textbf{ToDo: }#1}}{\hide{#1}}}
\newcommand\note[1]{\ifthenelse{\boolean{show_notes}}{\textcolor{blue}{#1}}{\hide{#1}}}
\newcommand\jochen[1]{\ifthenelse{\boolean{show_feedback}}{\textcolor{orange}{\textbf{Jochen: }#1}}{\hide{#1}}}
\newcommand\sebastian[1]{\ifthenelse{\boolean{show_feedback}}{\textcolor{orange}{\textbf{Sebastian: }#1}}{\hide{#1}}}
\newcommand\leo[1]{\ifthenelse{\boolean{show_feedback}}{\textcolor{orange}{\textbf{Leo: }#1}}{\hide{#1}}}

\newcommand\contrib[1]{\ifthenelse{\boolean{show_contributions}}{\textcolor{blue}{\textbf{Contributions:} #1}}{\hide{#1}}}

\newtheorem{definition}{Definition}[section]

\title{The elements of flexibility\\for task-performing systems}

\author{
  Sebastian Mayer\\
  Fraunhofer SCAI\\
  \texttt{sebastian.mayer@scai.fraunhofer.de}
  \And
  Leo Francoso Dal Piccol Sotto\\
  Fraunhofer SCAI\\
  \texttt{leo.francoso.dal.piccol.sotto@scai-extern.fraunhofer.de}
  \And
  Jochen Garcke \\
 Universität Bonn\\  Fraunhofer Center for Machine Learning and SCAI\\
  \texttt{garcke@ins.uni-bonn.de} 
}

\begin{document}
\twocolumn[
\maketitle
\begin{abstract}
What makes living systems flexible so that they can react quickly and adapt easily to changing environments? This question has not only engaged biologists for decades but is also of great interest to computer scientists and engineers who seek inspiration from nature to increase the flexibility of task-performing systems such as machine learning systems, robots, or manufacturing systems. In this paper, we give a broad overview of design features of living systems that are known to promote flexibility. We call these design features the ``elements of flexibility''. Moreover, to facilitate interdisciplinary, bio-inspired research that brings the elements of flexibility to man-made task-performing systems, we introduce a general formalism for system flexibility optimization. The formalism is intended to (i) provide a common language to communicate ideas about system flexibility among researchers with different backgrounds, (ii) help to understand and compare existing research on system flexibility, e.g., in transfer learning or manufacturing flexibility, and (iii) provide a basis for a general theory of system flexibility optimization. 
\end{abstract}
\vspace{1cm}
]
\clearpage

\twocolumn[
\tableofcontents
]
\clearpage

\nopagebreak

\section{Introduction}

\subsection{A general formalism for system flexibility optimization}

What makes systems flexible enough to handle a variety of tasks and easily adapt from one task to another? This question has engaged researchers for decades in various areas of computer science and engineering such as machine learning~\cite{thrun1998learning}, evolutionary computation~\cite{branke2012evolutionary}, robotics~\cite{mowforth1987ai}, and computer-integrated manufacturing~\cite{sethi1990flexibility}. Due to the very different types of tasks that are usually studied in these areas (e.g., classification tasks in machine learning and production tasks in computer-integrated manufacturing), research on system flexibility has taken place largely independently of each other. However, there is a substantial conceptual overlap and in recent years, the advancing digitization and ever more powerful computers also lead to increasingly overlapping practical flexibility matters. A good example are cyber-physical production systems that need more flexible machine learning systems that can learn process models based on few observations in order to increase the self-adaption capabilities to changing product requirements~\cite{monostori2016cyber}. Hence, there is an increasing demand for interdisciplinary research on system flexibility at the intersection of computer science and engineering.

To facilitate such interdisciplinary research, we propose in Section~\ref{sec:flexibility} a \emph{general formalism} to formulate and study flexibility problems for systems that are supposed to self-adapt to a variety of changing tasks. Based on the general, abstract notion of \emph{task-performing system}, we discuss how flexibility problems can be cast as optimization problems where adaption and reconfiguration cost are to be minimized over a space of system designs. The formalism is inspired by established formalisms from reinforcement learning~\cite{sutton2018reinforcement} and information-based complexity~\cite{packel1987information,novak2008tractability}. Beside being a useful working tool that provides a common language to communicate ideas among researchers with different backgrounds, we hope that the formalism also helps to understand and compare existing research on system flexibility from different areas in computer science and engineering. In Appendix~\ref{sec:boolean_circuit_evolvability}, we provide a complete example for an application of the formalism.

\subsection{The elements of flexibility}

The formalism that we propose in Section~\ref{sec:flexibility} provides a basis for a general theory of system flexibility. A central question that such a theory must try to answer is if there are generic design features of task-performing systems that facilitate flexibility and if so, what these design features are. We do not treat this question rigorously in this paper, but collect evidence for a number of universal design features by considering living systems.

Living systems are formidable natural examples of flexible, self-adapting task-performing systems. In living systems,  production, transportation and information-processing tasks are performed on every level of biological organization---from a single cell up to networks of organisms~\cite{alberts2002cell,palsson2015systems}. Moreover, learning occurs in various forms on every level of biological organization~\cite{tagkopoulos2008predictive,slijepcevic2021principles}. One can find myriads of fascinating examples of self-adaption cascades that result in flexibility within the lifespan of organisms~\cite{whitman2009phenotypic} or across generations through evolution~\cite{carroll2004dna}.

We conducted an extensive survey of biological literature, searching for design features that have been reported to promote flexibility. As a result, we have identified six elements that appear in all kinds of biological systems on all levels of biological organization:
\begin{enumerate}
    \item Hierarchy,
    \item Modularity,
    \item Weak regulatory linkage,
    \item Exploration,
    \item Degeneracy and neutrality,
    \item Weak links.
\end{enumerate}
We call these design features the \emph{elements of flexibility}. Remarkably, these elements play an essential role for both physical and computational tasks performed by biological systems. Some of the elements, like modularity, are well-established in the design of technical systems. Others, like weak links, have hardly been considered explicitly. To our knowledge, this work is the first to explicitly discuss all these elements from a general viewpoint of system flexibility. We discuss the elements of flexibility in detail in Section~\ref{sec:elements}.

The elements of flexibility are not merely interesting from a theoretical point of view, but we expect that an in-depth and holistic considerations of the elements can contribute to an improved flexibility of concrete task-performing systems, e.g., deep learning system or cyber-physical production systems. In this regard, Section~\ref{sec:elements} can serve as a starting point and source of inspiration for bio-inspired research on system flexibility. In Section~\ref{sec:outlook}, we sketch some general research directions based on the elements of flexibility.

\paragraph{Outline.}
This paper is intended for a multidisciplinary audience and organized as follows. In Section~\ref{sec:examples}, we give a brief introduction to existing flexibility research in computer-integrated manufacturing, machine learning, and biology. Then, in Section~\ref{sec:flexibility}, we introduce general concepts and a formalism for system flexibility optimization. Section~\ref{sec:elements} is devoted to the elements of flexibility. Finally, in Section~\ref{sec:outlook}, we give an outlook on a bio-inspired, interdisciplinary research agenda based on the elements of flexibility. In Appendix~\ref{sec:boolean_circuit_evolvability}, the interested reader finds an example how to apply the formalism to study the flexibility of logic circuits, which are popular model systems in theoretical computer science and computational biology. The presented example is also a simple demonstration of how hierarchy and modularity can promote flexibility.

\section{Existing research on system flexibility}

\label{sec:examples}

\begin{table}
  \centering
  \begin{tabular}{ll}
    \toprule
    \multicolumn{2}{l}{\textbf{Material tasks}}\\
    \textit{Type} & \textit{Goal}\\
    \midrule
    Production     & Change physical or chemical  \\
                   & properties of material \\
                   & to bring it in a target state\\
    \midrule
    Transportation & Change the location of material \\
    \midrule
    \multicolumn{2}{l}{\textbf{Immaterial tasks}}\\
    \textit{Type} & \textit{Goal}\\
    \midrule
    Control/       & Steer a system towards a desired,\\
    regulation     & state, e.g., using feedback loops\\
    \midrule
    Prediction and & Use virtual artifacts to estimate\\
    simulation     & the behavior of objects and\\
                   & processes\\
    \midrule
    Problem-solving & Use computational resources to\\
                    & find the solution to a problem\\
    \midrule
    Learning & Do the same task or tasks drawn\\
            & from the same population more\\
            & efficiently and effectively\\
            & the next time~\cite{simon1983should}\\
    \bottomrule
  \end{tabular}
  \vspace*{6pt}
  \caption{General high-level categories of tasks.}
  \label{tab:task_categories}
\end{table}

In this paper, we are generally interested in the flexibility of systems that perform tasks. Most abstractly, we can distinguish tasks into material (physical) tasks and immaterial (computational) tasks. In material tasks, the goal is to manipulate and transform physical objects in the environment. In immaterial tasks, the goal is to manipulate and transform data. Data can be considered as virtual objects in the environment. Note that an immaterial task necessarily requires a physical task to be performed. However, it is not the physical result that matters but the interpretation of the result as information. In Table~\ref{tab:task_categories}, we give a more fine-grained categorization of typical types of tasks. Now, a general, high-level definition of system flexibility for task-performing systems can be formulated as follows.

\begin{definition}
\label{defi:flexibility}
The flexibility of a task-performing system refers to its ability
\begin{enumerate}[label=(\roman*)]
    \item to easily adapt from being good at one task to being good at a related task, and
    \item to cope with a diversity of different tasks.
\end{enumerate}
\end{definition}

We formalize this definition and the involved concepts later in Section~\ref{sec:flexibility}. System flexibility in accordance with Definition~\ref{defi:flexibility} has been studied for specific classes of task-performing systems in various areas of computer science and engineering. We take a closer look at computer-integrated manufacturing and machine learning in Section~\ref{sec:manufacturing_flexibility} and Section~\ref{sec:learning_to_learn} since there are well-established subfields addressing system flexibility. Further areas where system flexibility has been studied include evolutionary computation~\cite{branke2012evolutionary}, robotics~\cite{mowforth1987ai}, supply chain management~\cite{tiwari2015supply}, and organization management~\cite{o2004ambidextrous}. To provide the necessary background for Section~\ref{sec:elements}, we give a brief introduction to system flexibility consideration in biology in Section~\ref{sec:evolvability}. We also briefly discuss how flexibility considerations in these areas can be connected.

\subsection{Flexibility in computer-integrated manufacturing}
\label{sec:manufacturing_flexibility}

\begin{figure}[t]
    \centering
\begin{tikzpicture}[
    scale=5,
    axis/.style={very thick, ->, >=stealth'},
    important line/.style={thick},
    dashed line/.style={dashed, thin},
    pile/.style={thick, ->, >=stealth', shorten <=2pt, shorten
    >=2pt},
    every node/.style={color=black}
    ]
    % axis
    \draw[axis] (0,0)  -- (1.1,0);
    \node(xline)[below] at (0.5,0) {Product variety};
    \draw[axis] (0,0) -- (0,0.8);
    \node(yline)[above, rotate=90] at (0,0.4) {Volume per variant};
    % Lines
    \path[draw,pil]
        (1,0.1) 
        -- (0.9,0.1) circle (0.3pt) node[above] {\tiny 1850}
        .. controls (0.5,0.1) and (0.3, 0.1) .. (0.2,0.3) circle (0.3pt) node[right] {\tiny 1913}
        .. controls (0.1,0.5) and (0.1,0.65) .. (0.15,0.7) circle(0.3pt) node[above] {\tiny 1955}
        .. controls (0.15,0.75) and (0.3,0.7).. (0.5,0.6) circle(0.3pt) node[below] {\tiny 1980}
        -- (0.75,0.45) circle(0.3pt) node[below] {\tiny 2000}
        .. controls (0.8,0.42) and (0.9,0.35) .. (1.08,0.2);
    \node[rotate=-10] at (0.55,0.17) {\small Craft production};
    \node[rotate=-5] at (0.25,0.8) {\small Mass production};
    \node[text width=1.7cm, align=center, rotate=-30] at (0.7,0.6) {\small Mass\\customization};
    \node[text width=1.7cm, align=center, rotate=-40] at (1.1,0.3) {\small Personalized production};
\end{tikzpicture}
    \caption{Development of production paradigms. Adapted from~\cite{koren2010global}.}
    \label{fig:production_paradigms}
\end{figure}
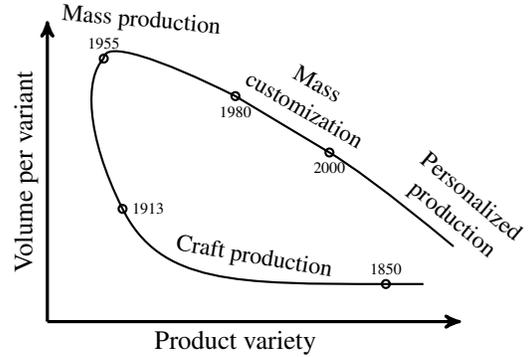

Flexibility is one of the outstanding characteristics of human beings. Consequently, in pre-industrial times, craft production mainly based on human labor was flexible. Mechanization and the advent of mass production brought efficiency to manufacturing, at the expense of flexibility, see Figure~\ref{fig:production_paradigms}. The extreme form of opting for efficiency is the assembly line, a manufacturing system that consists of a predefined sequence of machines and that is designed to perform a small number of very similar processes. As Herbert Simon, one of the pioneers in manufacturing flexibility and artificial intelligence, noted, ``mechanization has more often proceeded by eliminating the need for human flexibility---replacing rough terrain with a smooth environment---than by imitating it''~\cite{simon1960new}.  

Not all manufacturing work is amendable to the principles of mass production. Particularly in a business-to-business context, flexible, small to mid batch-size production has always remained necessary. The associated processes cannot be performed by an assembly line but require job shops, which are manufacturing systems that use functionally grouped general purpose equipment through which different work pieces can be flexibly routed. Job shops offer high flexibility at the expense of lower efficiency and costly machinery. Despite all the progress in automation technology, their operation typically requires highly skilled craft employees to the present day. It thus remains a challenge to increase manufacturing flexibility through innovation in automation technology~\cite{koren2010global}.

There are a number of established flexibility notions in computer-integrated manufacturing which address aspects of high- to low-level manufacturing tasks~\cite{browne1984classification,sethi1990flexibility,elmaraghy2005flexible}. On a higher level, referring to variations in amount and design of products, we have the following notions: 
\begin{enumerate}[label=H\arabic*.]
    \item \emph{Volume flexibility} refers to the different amounts of a part that can be produced profitably.
    \item \emph{Production flexibility} refers to the diversity of part types a manufacturing system can produce without major equipment changes.
    \item \emph{Product flexibility} refers to the ease of adding new parts to the existing part mix.
    \item \emph{Market flexibility} refers to the ease with which adaptions to changing market environments can be realized.
\end{enumerate}
On a lower level, referring to basic operations of manufacturing systems and their recombinability, we have:
\begin{enumerate}[label=L\arabic*.]
    \item \emph{Machine flexibility} refers to the diversity of operations a machine can perform and the ease with which the setup of a machine can be changed to produce a desired set of part types.
    \item \emph{Material flexibility} refers to the diversity of loading and unloading and transport situations a material handling system can cope with.
    \item \emph{Process flexibility} refers to the number of different processes a system can perform without major setups between process changeovers.
    \item \emph{Expansion flexibility} refers to the ease with which operations can be added to the manufacturing system.
\end{enumerate}

There are two further notions that do not directly address variations in tasks but robustness and fault-tolerance of the manufacturing system:
\begin{enumerate}[label=D\arabic*.]
    \item \emph{Operation flexibility} refers to the number of different ways in which a part can be produced. 
    \item \emph{Routing flexibility} refers to the number of routes through the system that lead to the same production outcome.
\end{enumerate}
However, robustness and flexibility are related, see Section~\ref{sec:related}. In particular, there is a strong relation to the element of flexibility ``degeneracy'', see Section~\ref{sec:degeneracy}.

For an in-depth discussion of these flexibility notions and their operationalization, we refer to~\cite{sethi1990flexibility,jain2013review}. Note that the systematization of the flexibility notions is still an ongoing research topic~\cite{jain2013review,perez2016review}. The unifying framework introduced in Section~\ref{sec:flexibility} can serve as a theoretical basis for such a systematization.

The aspects of manufacturing systems and manufacturing tasks that have been subject to flexibility considerations have always been highly influenced by progress in IT technology, see Table~\ref{tab:maturing} for an overview. Flexible manufacturing systems (FMS) focused on maximal flexibility with regard to tool operations and are nowadays standard in job shops~\cite{elmaraghy2005flexible}. Reconfigurable manufacturing systems (RMS) focused on reusable, modularized hard- and software components~\cite{koren1999reconfigurable}. RMS mark a milestone in the development of structurally flexible manufacturing systems. However, the concept was ahead of its time since computational and information-processing capacities were missing to handle the high reconfigurability in an automated way.

The emerging cyber-physical production systems (CPPS)~\cite{monostori2016cyber} promise to overcome these limitations. CPPS have computing, sensing and information processing capabilities in every physical component and are connected to cloud services by default. In consequence, there is access to an unprecedented amount of computation power and data about system states and environmental conditions in CPPS. There are high expectations that this allows to implement advanced self-adaption mechanisms in many components of CPPS which will in turn provide the flexibility necessary for a ``batch size one production''. In such a production regime, highly customized products can be produced with the efficiency of mass production~\cite{monostori2016cyber,lasi2014industry}. Machine learning is a key enabler for advanced self-adaption, amongst others, because it allows to deal with uncertainties, search more efficiently in large configuration spaces, and make automated decisions based on complex environmental cues~\cite{jamshidi2019machine,gheibi2021applying}.

Realizing the potential of CPPS is a long term goal that comes with many research challenges, see~\cite{monostori2016cyber} and the references therein. Concerning self-adaption and machine learning, challenges include learning efficiency and the handling of sudden or unknown changes in non-static, open-world environments~\cite{gheibi2021applying}. Moreover, many self-adaption mechanisms have to work in coordinated, cooperative ways within and across several system layers~\cite{muccini2016self}. Thereby it can easily happen that a change triggered by one adaption mechanism triggers a whole cascade of further adaptions and may even require an alteration in a subsequent adaption mechanism and its learned model~\cite{casimiro2021self}. For all these reasons, the learning systems used within adaption mechanisms should themselves be flexible, easily adapting learned models to changed circumstances using only little extra data. Self-adaption and CPPS are thus not merely application areas for machine learning but drivers for research in \emph{meta-learning} and \emph{learning to learn} which address the flexibility of learning systems, see Section~\ref{sec:learning_to_learn}. Moreover, flexibility considerations are becoming more complex, taking very deep aspects of the production processes into account such as material properties and not only the operation phase but also development and ramp-up. Hence, software systems such as CAD, CAE, and CAM are increasingly included in automation and flexibility considerations~\cite{ganser2022knowledge}. 

\begin{table}[t]
  \centering
  \begin{tabular}{lll}
    \toprule
    \textbf{Decade} & \textbf{Concept} & \textbf{Maturing IT technology} \\
    \midrule
    1970s & FMS     & microprocessors, \\
          &         & embedded systems, \\
          &         & procedural programming \\
    \midrule
    1990s & RMS     & modular programming,\\
          &         & open architectures \\
    \midrule
    Since & CPPS    & digital sensory equipment,\\
    2010s &         & wireless communication,\\
          &         & big data, cloud computing,\\
          &         & artificial intelligence/\\
          &         & machine learning\\
    \bottomrule
  \end{tabular}
  \vspace*{6pt}
  \caption{IT drivers of manufacturing flexibility. The table lists IT technologies that are or were becoming sufficiently mature or broadly available in the corresponding time period and are or were included in manufacturing system concepts for flexibilization.}
  \label{tab:maturing}
\end{table}

\subsection{Flexibility in machine learning}
\label{sec:learning_to_learn}

The flexibility of human learning is one of the most prominent examples of system flexibility. Humans are able to learn myriads of different tasks, e.g., language understanding, object recognition, motor skills and concept formation. Usually, we learn these tasks not apart from each other but by being exposed to a continual stream of diverse tasks. In doing so, an outstanding characteristic of human learning is that we can generalize correctly from extremely few examples---often just a single example suffices to learn a new task~\cite{lake2017building, ahn1993psychological}.

In artificial intelligence, it is a long-standing goal to build learning systems that achieve the same learning flexibility as human beings. While classical machine learning focuses on learning systems that improve their performance at one given task with training experience, \emph{learning to learn} or \emph{meta-learning}~\cite{thrun1998learning,vilalta2002perspective} aims for flexible learning systems that improve their performance at each task both with training experience and with the number of tasks.

At the heart of every learning system is a statistical model. In classical machine learning, the model is chosen by the system designer to reflect general structural knowledge about the given underlying task. Then, a learning algorithm is used to fit the model's free parameters to accurately capture the patterns in the training data. With learning to learn, the knowledge still imposed by the system designer becomes increasingly abstract and unspecific. It is the learning system itself that must be able to capture and represent causal relationships and high-level structures and the patterns to be recognized are relations between the different learning tasks. Learning to learn can thus be understood as a shift of focus, from classical pattern recognition to representation learning and automated model building~\cite{lake2017building}. The source of learning efficiency gains in the face of new tasks are then proper ways of transferring parts of the learned representations and model parameters between related learning tasks, which is called \emph{transfer learning}~\cite{pan2010transfer, yang2020transfer}. Note that this is closely related to \emph{few-shot learning}~\cite{wang2020few} which studies how proper representations of prior knowledge help machine learning models to generalize well to new, related tasks using only few samples.

Flexibility of learning systems is highly relevant for manufacturing flexibility for several reasons.The recent success stories in machine learning are based on tremendous amounts of training data or relatively cheap simulations~\cite{von2021informed}. In manufacturing contexts, labelled data is often scarce and expensive to create. Simulations have to take into account more physial effects and are thus more expensive. Hence, to fully leverage the learning potential in CPPS, the machine learning systems used within CPPS have to be flexible, easily adapting to new tasks using only few examples. Moreover, meta-learning techniques provide the basis for automated machine learning (AutoML) tools~\cite{karmaker2021automl}. These tools automate steps in the machine learning pipeline such that application domain experts can more easily experiment with machine learning and thereby increase the development speed of machine learning for production. 

\subsection{Flexibility in living systems}
\label{sec:evolvability}

Organisms consist of cells as basic construction units and each cell has a copy of the genome. The genome is the totality of all heritable information about how to construct and operate the organism and all of its components and processes. The concretely given heritable information of an individual organism is called its \emph{genotype}. The sum of all observable characteristics and traits of the organism is called the \emph{phenotype}.

The ultimate purpose of an organism is to reproduce to guarantee the survival of its lineage. For this, the organism has to survive long enough in an environment for which it has to perform tasks like identifying food, moving around, seeking shelter etc. Within an organism, we find myriads of subsystems that perform tasks, e.g., metabolic pathways that process nutrients, or sensory organs like eyes and ears that process signals from the environment. If the environment changes, many goals will be the same but the way in which an organism can achieve the goals may differ. The changes in the environment may also be drastic such that new tasks appear for which novel traits are necessary.

One can understand all this as organisms being constantly in a process of finding and applying solutions to tasks posed by environments~\cite{lewontin2001triple}. In contrast to classical problem-solving, the solution object (the organism) is already there, the result of solving a sequence of previously posed tasks. If there is a new task, nature cannot start from scratch but the solution object can only be adapted or adapt itself. Consequently, it is advantageous for organisms to be flexible in the sense of Definition~\ref{defi:flexibility}. Adaption can basically take place in two forms: within the lifespan of an organism ,which is collectively referred to as \emph{phenotypic plasticity}~\cite{whitman2009phenotypic}, or across generations through natural evolution, which is called \emph{evolutionary adaption}.

\begin{figure}[t]
    \centering
    \begin{tikzpicture}[node distance=1cm]
    \tikzstyle{every state}=[fill=white,draw=black,text=black]
        \node[punkt, minimum width=3.5cm, minimum height=1.5cm, text width=3.3cm, text depth=1.3cm, align=left] (organism) {Organism};
        \node[punkt] (genome) at (0,0) {Genome};
        \node[text width=1.2cm, align=left, left=0.3cm of organism] (mutation) {\small Random mutation};
        \node[text width=1.5cm, right=0.5cm of organism] (variation) {\small Useful phenotypic variation};
        \path (mutation.east) edge[pil] (genome.west)
              (organism.east) edge[pil] (variation.west);
    \end{tikzpicture}
    \caption{The organism plays an active role in the evolutionary process by rendering random genetic change into potentially useful phenotypic variation.}
    \label{fig:evolvability}
\end{figure}
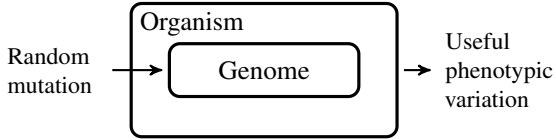

Evolutionary adaption of organisms to changing environmental conditions happens through a population-based mutation-variation-selection cycle. Mutations introduce random changes to genomes which leads to phenotypic variation among the organisms in the population such that their fitness to the environment becomes different. Individual organisms with a higher fitness have a better chance to survive and reproduce such that they eventually will dominate the population. Nature ``selects'' the fittest. Though mutations are random, the effects on the phenotype are not. Biologists have made the striking observation that the diversity in the genetic material of different organisms is much smaller than the diversity in their anatomy, morphology, and physiology. Moreover, the complexity of observable phenotypic adaptions is way to large for a one-to-one relationship between genetic and phenotypic change. In consequence, the organism must play the role of a transformation engine that links genetic change in a complex, non-linear way to phenotypic change, see Figure~\ref{fig:evolvability}. The organisms capacity to render small, random genetic changes into potentially useful and novel phenotypic traits is called its \emph{evolvability}~\cite{kirschner1998evolvability}. Higher evolvability means higher flexibility in the sense of Definition~\ref{defi:flexibility}. The theory of facilitated variation~\cite{gerhart2007theory} is devoted to explaining what properties of organismal design and biological processes increase evolvability. This theory has been one of our starting points to elaborate the elements of flexibility as presented in Section~\ref{sec:elements}.

In contrast to evolutionary adaption, phenotypic plasticity refers to adjustments of an organism to environmental conditions within its lifespan, which do not the result from heritable changes to the genome. Hence the name ''phenotypic plasticity```as the same genotype can result in different phenotypes. Phenotypic plasticity can be observed with regard to all aspects of an organism: its form and structure (morphology); the way chemical and physical functions are carried out in organs, cells, and biomolecules (physiology); its behavior, i.e., internally computed and coordinated responses to various stimuli or inputs. Phenotypic plasticity can refer to developmental effects, which lead to long-lasting adjustments to environmental conditions, but also includes acclimatization, i.e., temporary, reversible adjustment that happens in short periods of time (hours to weeks). Learning in the classical sense is phenotypic plasticity that changes behavior.

Remarkably, there is a strong overlap between the design characteristics of biological systems and processes that are considered to be a source of phenotypic plasticity~\cite{whitman2009phenotypic,price2003flexibility} and the ones that are considered to increase evolvability. Moreover, it is particularly remarkable that conservation of certain core functionalities, modular and hierarchical organization, and the use of exploration, which play an important role in evolvability~\cite{kirschner1998evolvability,gerhart2007theory}, are also essential ingredients for the flexibility of human learning and thinking~\cite{lake2017building,butz2016mind}. The omnipresence and universality of certain design characteristics that foster flexibility in living systems with regard to very different tasks is what makes us believe that their role should also be investigated more deeply and holistically for man-made task-performing systems. Promoting this idea and research related to this idea is the main motivation for introducing the elements of flexibility in Section~\ref{sec:elements}.

\section{System flexibility optimization}
\label{sec:flexibility}

\begin{figure*}[ht]
    \centering
    \begin{tikzpicture}[node distance=0.5cm]
    \tikzstyle{every state}=[fill=white,draw=black,text=black]
        \node[punkt] (tps) {\small Task-performing system};
            \node[punkt, below=of tps] (conf) {\small Configuration};
            \node[punkt, right=of conf] (design) {\small Design};
            \node[punkt, left=of conf] (ls) {\small Learning system};
            \path[draw,pil] (tps.south) --++(0,-0.15cm) -- ++(-3.06cm,0) -- (ls.north);
            \path[draw,pil] (tps.south) --++(0,-0.15cm) -- ++(3.06cm,0) -- (design.north);
            \path (tps.south) edge[pil] (conf.north);
        \node[punkt, above=of tps] (cs) {\small Cost function};
            \path (cs.south) edge[pil, dashed] (tps.north);
            \node[punkt, above right=0.5cm and -0.75cm of cs] (runtime) {\small Runtime};
            \node[punkt, above left=0.5cm and -0.75cm of cs] (adaption) {\small Adaption};
            \path[draw, pil, arrows = {-Stealth[fill=white, inset=0pt, angle=90:5pt]}] (adaption.south) -- ++(0,-0.2cm) -- ++(1.81cm,0) -- (cs.north);
            \path[draw, pil, arrows = {-Stealth[fill=white, inset=0pt, angle=90:5pt]}] (runtime.south) -- ++(0,-0.2cm) -- ++(-1.81cm,0) -- (cs.north);
                \node[punkt, above right=0.5cm and -1cm of adaption] (reconf) {\small Reconfiguration};
                \node[punkt, above left=0.5cm and -0.75cm of adaption] (learning) {\small Learning};
                \path[draw, pil, arrows = {-Stealth[fill=white, inset=0pt, angle=90:5pt]}] (learning.south) -- ++(0,-0.2cm) -- ++(1.81cm,0) -- (adaption.north);
                \path[draw, pil, arrows = {-Stealth[fill=white, inset=0pt, angle=90:5pt]}] (reconf.south) -- ++(0,-0.2cm) -- ++(-1.565cm,0) -- (adaption.north);
        \node[punkt, right=3cm of tps] (task) {\small Task};
            \path (tps.east) edge[pil, dashed] (task.west);
            \node[punkt, right=1cm of task] (pm) {\small Performance measure};
            \node[punkt, above=of pm] (cons) {\small Constraints};
            \node[punkt, above=of cons] (goal) {\small Goal};
            \node[punkt, above=of goal] (env) {\small Environment};
            \path (task.east) edge[pil] (pm.west);
            \path[draw,pil] (task.east) -- ++(0.5cm,0) -- ++(0,1.3cm) -- (cons.west);
            \path[draw,pil] (task.east) -- ++(0.5cm,0) -- ++(0,2.55cm) -- (goal.west);
            \path[draw,pil] (task.east) -- ++(0.5cm,0) -- ++(0,3.8cm) -- (env.west);
        \node[punkt, above=of task] (tc) {\small Task context};
            \path (tc.south) edge[pil] (task.north);
        \node[punkt, above=of tc] (fm) {\small Flexibility measure};
            \path (fm.south) edge[pil, dashed] (tc.north);
            \path[draw, pil, dashed] (fm.west) -- ++(-0.25cm,0) -- ++(0,-1.65cm) -- ++(-4.05,0) -- (tps.north);
            \path[draw,pil] (fm.west) -- ++(-0.5cm,0) -- ++(0,-1.2cm) --  (cs.east);
        \path[draw, pil, dashed] (7.5,-1.5) -- (8.5,-1.5) node[right] {\small is applied to};
        \path[draw, pil] (7.5,-2) -- (8.5,-2) node[right] {\small has a};
        \path[draw, pil, arrows = {-Stealth[fill=white, inset=0pt, angle=90:5pt]}] (7.5,-2.5) -- (8.5,-2.5) node[right] {\small specialization of};
    \end{tikzpicture}
    \caption{General concepts involved in system flexibility optimization.}
    \label{fig:flexibility_concepts}
\end{figure*}
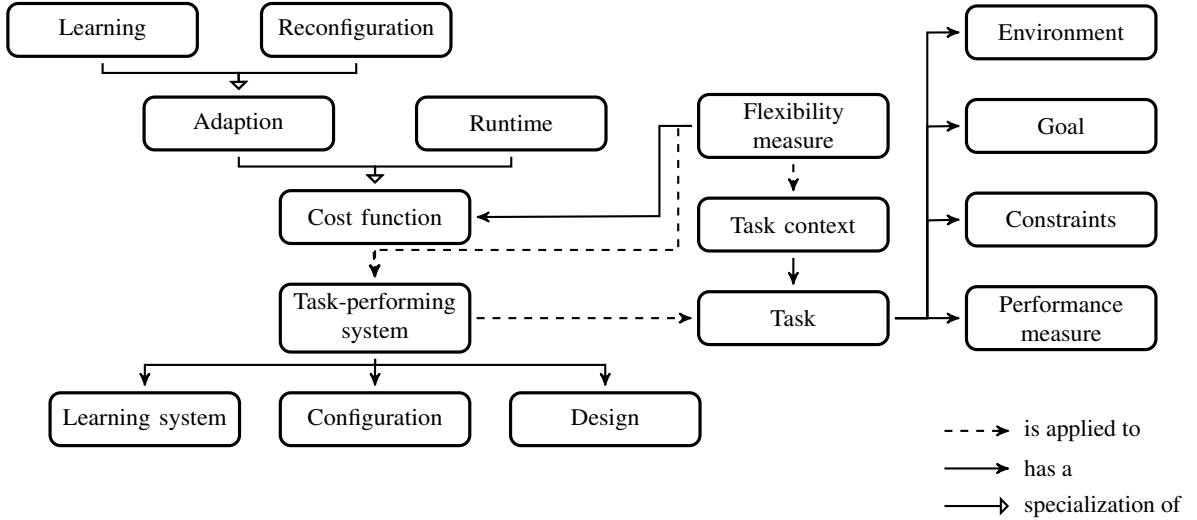

In this section, we take some first steps towards a general, formal theory of system flexibility that equally applies to any area in which systems appear that have to perform tasks.  

The key idea is that we cast the problem of improving system flexibility such as informally defined in Definition~\ref{defi:flexibility} as an optimization problem over a space of task-performing systems. For this, we first have to specify a \emph{task context}, formally given by a tuple
$$
W=(\mathcal T, Q),
$$
where~$\mathcal T$ is a set of tasks and $Q$ a (transition) probability distribution on~$\mathcal T$. The distribution $Q$ describes which tasks and which order of tasks the system is likely to see. We can also consider this as modelling the importance of a task, where more frequently appearing tasks are more significant. Next, we have to specify a \emph{flexibility measure}~$f_W(S)$ which quantifies in a suitable way how flexible a given task-performing system~$S$ is in the given task context~$W$. Larger values of the flexibility measure indicate larger flexibility. There is more than one way to formalize the two flexibility aspect ``adaptability'' and ``task diversity'' which we have informally described in Definition~\ref{defi:flexibility}. We elaborate some flexibility measures in Section~\ref{sec:flexibility_notions}. Let us only mention here that these flexibility measures are based on certain \emph{cost functions} that measure the cost to reconfigure or adapt a system. Maximizing flexibility then amounts to \emph{minimizing reconfiguration or adaption cost} in a suitable way. 

Abstractly, we may now define the general system optimization problem as follows.
\begin{definition}\label{defi:flexibility_optimization}
Given a space of task-performing systems~$\mathcal S$, a task context $W=(T,Q)$, and a flexibility measure
$$
f_W: \mathcal S \to \mathbb{R}
$$
for the given task context,
the goal of system flexibility optimization is to find the task-performing system $S^*$ with the maximal flexibility in the given task context,
$$
 f_W(S^*) = \max_{S \in \mathcal S} f_W(S).
$$
\end{definition}

In practice, we will seldom be in a situation where we can perform an extensive search in a large space of different task-performing systems suitable for the considered task context. Hence, we will rarely have the chance to find the optimal system. Rather, we will be interested to systematically understand which characteristics in the \emph{system design} allow us to increase the value of the flexibility measure~$f_W$. Here the elements of flexibility, which we introduce in Section~\ref{sec:elements}, come into play. The basic hypothesis driving this paper is that the elements of flexibility, properly incorporated in the design of a task-performing system, allow to attain high flexibility values for various flexibility measures in many task contexts. 

In the following, we discuss the general concepts that have to be put to concrete terms in order to approach system flexibility optimization in a systematic way. In this way, we provide a formalism that is supposed to help interdisciplinary teams in communicating about system flexibility. For an overview of the involved concepts, see Figure~\ref{fig:flexibility_concepts}. To demonstrate the application of the formalism, we discuss an example from computational biology in the appendix, see Section~\ref{sec:boolean_circuit_evolvability}.

%New Cybernetics studies systems that perform tasks. This includes technical systems such as computers and machines, biological systems such as metabolic pathways, but also immaterial systems such as artificial neural networks which are simulated on a computer. In particular, New Cybernetics is interested in \emph{intelligent agents}, i.e., systems that act autonomously, directing its activity towards achieving goals (i.e. it is an agent), upon an environment using observation through sensors and consequent actuators (i.e. it is intelligent). Technical examples of intelligent agents include robots and other cyber-physical systems, examples for biological agents are single cells or whole 
%organisms.

%In this section, we collect the basic concepts and aspects of task-performing systems which we need to discuss system flexibility from a general point of view.

%When we speak of \emph{objects} in the following, then we mean both physical objects and virtual objects (i.e. data).

\subsection{Tasks}
\label{sec:task}

A key ingredient of system flexibility optimization is the definition of a task context. It describes the tasks with regard to which we wish to have a flexible system. A comprehensive formulation of a task includes
\begin{enumerate}[label=\roman*)]
 \item a description of the \emph{goal},
 \item a description of \emph{additional requirements or conditions},
 \item a description of the \emph{environment},
 \item and a \emph{performance measure} indicating if the task is performed sufficiently well or quantifying how good a system is in performing the task, e.g. in terms of quality or resource consumption.
\end{enumerate}

Let us explain why it is important to include all these aspects in the task definition. When speaking informally about a task, we usually associate the task with the goal to be achieved. However, additional requirements, environmental conditions and how we measure performance affect the character of a task and its difficulty. Hence, properly formulated, performing a task means to achieve a goal in a given environment according to certain performance criteria while possibly fulfilling additional requirements or conditions. This is easily seen by an illustrative example. Consider a task where the goal is throw a ball into a basket. An example for a further requirement is from where on the court the ball has to be thrown into the basket. Examples for environmental conditions are wind direction and strength. The performance measure can simply be the percentage of throws where the ball hits the basket but we can also be more demanding and only count those throws as hits where the ball does not touch the rim of the basket. Everybody who has ever trained to throw a ball into a basket will remember how changing the position, changing wind conditions, and demanding a cleaner throw requires extra training because one has to learn how these different scenarios affect how to exactly perform the movement of the arm. Hence, when we speak of ``throwing a ball into a basket'' we may actually refer to a whole context of tasks that share the same goal but vary in the other aspects.

Formally, we can consider a task as a tuple
$$
T=(E,F,t,P)
$$
with the following elements. We are in an environment described by a space~$E$ of all possible environmental states. Additional requirements and conditions lead to a subset~$F \subseteq E$ of feasible states that any system that is trying to perform the task is allowed to set the environment to or has to try to keep the environment in. The goal is formally described by a map
$$
t: X \to Y,
$$
where $X \subset F$ is a subset of initial states and $Y \subset F^*$ a subset of target states or target state sequences ($F^*$ is the Kleene closure of $F$, i.e., the set of all vectors of arbitrary dimension with components in $F$). The map $t$ describes which target state or sequence of states is supposed to be reached from which initial state. Typically, the function~$t$ will map initial states to target states if the goal is about achieving certain static properties of objects in the environment, e.g., a target position. If the goal is about achieving certain dynamic properties, e.g., maximizing the speed of a moving object, then the target will often be described by a sequence of states that represent the desired dynamic property. We can think of the map~$t$ as an oracle that perfectly describes \emph{what} to achieve. That a system performs a task basically means that it runs a process that approximates the map $t$ sufficiently well. We describe this formally in Section~\ref{sec:performing_tasks}, after we have explained the notion of task-performing system in more detail. In Section~\ref{sec:performing_tasks}, we also give a formal description of the performance measure~$P$. Note that for many real-world tasks, it will be hard, impractical or even impossible to analytically specify the map~$t$ but we might approximately realize it by a high-fidelity simulation engine or have finitely many samples of it in the form of training and test data.  

\subsection{Task-performing systems}

In general, a task-performing system $S$ is a collection of components that serve three purposes:
\begin{itemize}
    \item Storage components hold collections of objects.
    \item Sensory components can perceive signals from the environments, e.g., cameras or eyes.
    \item Manipulatory components can take and manipulate objects in the environment of the system or in storage units, e.g.,  actuators, tools, computation units, enzymes and other functional proteins, etc.
\end{itemize}

\begin{figure}
    \centering
    \begin{tikzpicture}[node distance=1cm]
    \tikzstyle{every state}=[fill=white,draw=black,text=black]
        \node[punkt] (system) {Task-performing system~$S$};
        \node[punkt, below=of system] (environment) {Environment~$E$};
        \path[draw,pil] (system.east) -- ++(1cm,0) -- node[right] {Action} ++(0,-2.05cm) -- (environment.east);
        \path[draw,pil] (environment.west) -- ++(-1cm,0) -- node[left] {Percept} ++(0,2.05cm) -- (system.west);
    \end{tikzpicture}
    \caption{System-environment interaction.}
    \label{fig:system_environment_interaction}
\end{figure}
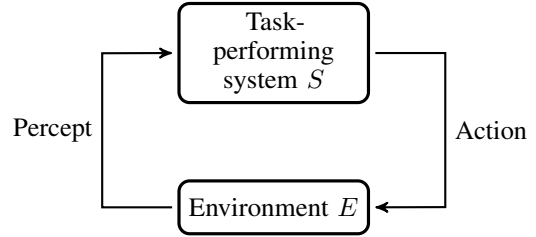

As in reinforcement learning, we do not consider the task-performing system as a part of the environment but that it receives signals/percepts from the environment and extracts and manipulates objects from the environment through sequences of actions, see Figure~\ref{fig:system_environment_interaction}. If the task-performing system is equipped with a learning system as described in Section~\ref{sec:learning_system}, then it is an \emph{intelligent agent}. For a best-practice discussion where to draw the boundaries between system and environment, we refer to~\cite[Chap. 3]{sutton2018reinforcement}.

\subsubsection{System design and configuration}

A \emph{system design} or \emph{system architecture} describes which components a system consists of, how they are arranged and what connections between the components exist or are possible. The system design determines which basic actions or operations can be executed by the system.

The components that make up the system can be configurable. A \emph{system configuration} is a concrete value assignment for the set of adjustable parameters of the system. The nature and complexity of the adjustable parameters can be quite different depending on the type of system under consideration. Moreover, which adjustable parameters we actually take into consideration will depend on the tasks that we are interested in. Examples for adjustable parameters are:
\begin{itemize}
    \item Simple numeric parameters of a controller.
    \item The weights and intercept terms in an artificial neural network.
    \item The program that is to be executed by a computer.
    \item Genes in the genome of an organism that are switched on or off by a natural mutation or in a controlled experiment.
\end{itemize}

Given a system $S$, we denote by $\Gamma(S)$ the \emph{system configuration space}, which is the space of all possible system configurations. Initially, any system $S$ has to be in some \emph{initial configuration} $\gamma_0 \in \Gamma(S)$.

\subsubsection{Algorithms and processes}

A \emph{process} is a finite sequence of actions that manipulate physical objects (e.g., substrates or work pieces) or virtual objects (i.e., data) in the system's environment. The object or environmental state that the first action manipulates is called the \emph{input}. The object or environmental state that results from the manipulation of the last action is called the \emph{output}.

An \emph{algorithm} is a unambiguous description of a sequence of instructions that tell the system which actions to perform. Processes result from executing algorithms on a system. For technical systems, the algorithm underlying a process is always known, while for biological systems, the algorithm that leads to the observed process might be unknown. If a task-performing system has no configurable components, then there is exactly one, hard-coded algorithm that is implemented by the system and the resulting processes have hard-coded characteristics. Configurable systems can implement more than one algorithm and a change in the system configuration can alter the characteristics of actions. Given a system $S$ and a configuration~$\gamma \in \Gamma(S)$, we denote by $S(\gamma)$ the algorithm that is implemented on the system when choosing the configuration~$\gamma$. The algorithm $S(\gamma)$ and the associated processes realize a certain input-output relationship, which is mathematically given by a function
$$
 f_{S(\gamma)}: \tilde X \to \tilde Y,
$$
where $\tilde X \subseteq E$ is the set of accepted inputs and $\tilde Y \subseteq E$ the set of realizable outputs. Running an algorithm on a system incurs costs, such as the required runtime or some energy consumption. These cost can generally depend on the input~$x \in \tilde X$.

% The relationship between algorithm and resulting process can be non-deterministic, i.e. in a given situation the algorithm might not always instruct the same but different actions.
%Only if the system does not interact with the environment but only acts upon it and instructions in the algorithm do not make use of randomness, then the process resulting from executing the algorithm will always be the same for the same input (except for random fluctuations due to noise). If the system is equipped with sensory elements that allow it to receive percepts from the environment, then the system can interact with the environment and the actually realized sequence of actions may depend on the sequence of received percepts. So the process resulting from executing an algorithm does not only depend on the input but the states of the environment.

\begin{definition}
The \textbf{execution cost}
$$
 c^{\mathrm{run}}(S, \gamma, x)
$$
are the cost of executing the algorithm~$S(\gamma)$ on the system~$S$ when the input is~$x \in \tilde X$. The \textbf{worst-case execution cost} associated to a configuration are given by
$$
 c^{\mathrm{wor-run}}(S, \gamma) := \max_{x \in \tilde X} c^{\mathrm{run}}(S, \gamma, x),
$$
the \textbf{best-case execution cost} associated to a configuration are given by
$$
 c^{\mathrm{best-run}}(S, \gamma) := \min_{x \in \tilde X} c^{\mathrm{run}}(S, \gamma, x),
$$
and given a probability distribution $p$ on the space of inputs $\tilde X$, the \textbf{average-case execution cost} associated to a configuration
are given by
$$
 c^{\mathrm{avg-run}}(S, \gamma) := \mathbb E \left[c^{\mathrm{run}}(S, \gamma, X)\right],
$$
where $X$ is random variable taking values in $\tilde X$ with distribution $p$.
\end{definition}

For simplicity, we consider only the worst-case execution cost in this paper and denote them by
$$
c^{\mathrm{run}}(S, \gamma) := c^{\mathrm{wor-run}}(S, \gamma).
$$

Note that since a configuration change can alter the characteristics of actions it may change the cost and peformance of an algorithm run.

\subsubsection{Performing a task}
\label{sec:performing_tasks}

Recall from Section~\ref{sec:task} that a task is formally given by a tuple~$T=(E,F,t,P)$. Given a task $T$, a system $S$ in a configuration $\gamma \in \Gamma(S)$ is able to perform the task if the actions in the processes resulting from the algorithm $S(\gamma)$ keep the environment within the set of feasible states $F$ and the algorithm~$S(\gamma)$  leads to an input-output relationship~$f_{S(\gamma)}$ that approximates the map
$$
    t: X \to Y
$$
sufficiently well according to some performance criteria. What ``sufficiently well'' means precisely is encoded by the performance measure~$P$. For instance, the performance measure~$P$ could quantify the distance between~$f_{S(\gamma)}$ and~$t$ according to a norm~$\|\cdot\|$,
$$
 P(S(\gamma)) = \|f_{S(\gamma)} - t\|.
$$
For simplicity, we assume for the remainder of this paper that $P$ is binary with
$$P(S(\gamma))=1$$
if the system  performs the task sufficiently well according to the performance criteria, and~$P(S(\gamma) )=0$ otherwise.

\subsubsection{Learning system}
\label{sec:learning_system}

If a task-performing system~$S$ should be able to put itself into the correct configuration for a task~$T$ and this configuration is not known in advance, then the task-performing system must be able to learn and to do problem-solving. To this end, it needs a subsystem which we generally refer to as the  \emph{learning system}. The learning system operates on the system configuration space~$\Gamma(S)$.

In the case of problem-solving, the learning system has a model of the task $T$ that is accurate and powerful enough to do a targeted search in the configuration space~$\Gamma(S)$ without further interaction with the environment. In the case of learning, the learning system needs observations and trial-and-error runs to direct the search in the configuration space and to learn a model of the task using some learning algorithm, e.g. backpropagation if the learning system is given by a artificial neural network.

Note that the task-performing system and the learning system can coincide. This is for instance the case when we consider the flexibility of a deep artificial neural network architecture with regard to a number of object recognition tasks.

\subsection{Flexibility measures}
\label{sec:flexibility_notions}

Recall the informal definition of system flexibility given in Definition~\ref{defi:flexibility}. It has two aspects. The first, which we call \emph{adaptability}, puts emphasis on how easy it is to switch between tasks. The second, which we call \emph{task diversity}, puts emphasis on the totality of different tasks that can be performed by the system. In this section, we discuss how to make these two aspects quantifiable such that we can optimize a system and its design in terms of these two aspects.

Concerning adaptability, we first consider the special case where the tasks and suitable system configurations for the tasks are already known. Then we are only concerned with the \emph{reconfigurability} of the systems, i.e., how costly it is two switch between configurations. Subsequently, we consider the general adaptability scenario where suitable configurations for tasks are not known in advance and the system has to do some form of problem-solving or learning to find suitable configurations. 

\subsubsection{Reconfigurability}

Consider a task-performing system $S$ with configuration space~$\Gamma(S)$ and two different tasks~$T_1$ and~$T_2$. Let us assume that for both tasks suitable configurations~$\gamma_1 \in \Gamma(S)$ and $\gamma_2 \in \Gamma(S)$ are already known that allow the system to perform task~$T_1$ and~$T_2$, respectively. Then no learning or problem-solving is required. However, although~$\gamma_1$ and~$\gamma_2$ are known, switching between these configurations can still require some operations that are not for free. For instance, a reconfiguration could involve to load a substantial amount of data from disk to memory or, in the case of a manufacturing system, it could require to change some tools that leads to a downtime of the system. We denote the cost (time, energy, number of operations, etc.) that are necessarily incurred when moving from one configuration to another as \emph{reconfiguration cost}. A first, basic formal definition can be given independently of the task concept as follows.

\begin{definition}
Given two configurations $\gamma_1, \gamma_2 \in \Gamma(S)$ of a system $S$, the \textbf{reconfiguration cost}
$$
 c^{\mathrm{reco}}(S, \gamma_1, \gamma_2)
$$
are the cost resulting from taking the necessary steps to get from configuration $\gamma_1$ to configuration $\gamma_2$.
\end{definition}

As we are interested in low reconfiguration cost for configurations that are associated to tasks, we also would like to have a definition of reconfiguration cost in terms of two tasks. For this, we have to take into account that there might be more than one suitable configuration per task. This leads us to the following formal definition.

\begin{definition}\label{defi:min_reco_cost}
Given two tasks $T_1$ and $T_2$ that can be performed by a system $S$, let $\Gamma(S,T_i)$ be the set of configurations in which system $S$ can perform task $T_i$. Then, the \textbf{minimal reconfiguration cost} to switch from task $T_1$ to task $T_2$ are formally given by
$$
 c^{\mathrm{min-reco}}(S, T_1, T_2) \\= \min_{\gamma_1,\gamma_2} c^{\mathrm{reco}}(S,\gamma_1, \gamma_2),
$$
where the minimum is taken over all $\gamma_1 \in \Gamma(S,T_1)$ and~$\gamma_2 \in \Gamma(S, T_2)$.
\end{definition}

Given two tasks, the smaller the reconfiguration cost for the corresponding system configurations, the more easily we can reconfigure the task-performing system. Hence, with regard to two tasks, it seems natural to define the \emph{reconfigurability} of a system either as the reciprocal of the minimal reconfiguration cost,
$$
\mathrm{reconfigurability} = \frac{1}{\mathrm{reconfiguration\ cost}}
$$
or as the negative reconfiguration cost
$$
\mathrm{reconfigurability} = -\mathrm{reconfiguration\ cost}.
$$
In both cases, smaller reconfiguration cost means higher reconfigurability. The latter approach is more suitable from a numerical point of view as taking a reciprocal can have numerical stability issues. Hence, we only use the latter in the following. Generalizing the reconfigurability notion to more than two tasks, there is more than one way to do this. One option is to take a conservative, worst-case point of view. Then we ask for the maximal pair-wise minimal reconfiguration cost that can occur for a given system in a tasks context.

\begin{definition}
The \textbf{worst-case reconfiguration cost} of a system $S$ in a tasks context $W=(\mathcal T, Q)$ are given by
    $$
     c_W^{\mathrm{wor-reco}}(S) := \max_{T_1, T_2 \in \mathcal T} c^{\mathrm{min-reco}}(S, T_1, T_2).
    $$
Then, we define the \textbf{worst-case reconfigurability} as
$$
 \rho_W^{\mathrm{wor}}(S) := -c_W^{\mathrm{wor-reco}}(S). 
$$
\end{definition}

Another, less conservative approach is to consider the minimal reconfiguration cost in the average-case.

\begin{definition}\label{defi:reconfigurability}
Given a task context~$W=(\mathcal T, Q)$, let $T_1$ and $T_2$ be two independent random tasks taking values in $\mathcal T$ with distribution $Q$. Then, the \textbf{average-case reconfiguration cost} of a system $S$ in the context~$W$ is given by 
    $$
     c_W^{\mathrm{avg-reco}}(S) := \mathbb E\left[c^{\mathrm{min-reco}}(S, T_1, T_2)\right].
    $$
Then, we define the \textbf{average-case reconfigurability} as
$$
 \rho_W^{\mathrm{avg}}(S) := -c_W^{\mathrm{avg-reco}}(S). 
$$
\end{definition}

Using $f_W = \rho_W^{\mathrm{wor}}$ or~$f_W = \rho_W^{\mathrm{avg}}$ as the flexibility measure in Definition~\ref{defi:flexibility_optimization}, we optimize the task-performing system for reconfigurability.  

\subsubsection{Adaptability}
\label{sec:adalearn}

If a suitable configuration for a new task~$T$ is not already known, then the learning system has to do some kind of problem-solving or learning. In addition to reconfiguration cost, adapting to the new task~$T$ then also incurs search or learning cost, e.g., cost to compute a suitable configuration, the number of observations, or trial-and-error runs to find a suitable reconfiguration. By \emph{adaption cost} we denote the combination of learning and reconfiguration cost. Formally, we define adaption cost as follows. 

\begin{definition}\label{defi:learning_cost}
Given a task $T$ and a system configuration~$\gamma \in \Gamma(S)$, the \textbf{adaption cost}
$$
 c^{\mathrm{ada}}(S,\gamma,T)
$$
denote 
the effort that is necessary for the system~$S$ in configuration $\gamma$ to find and attain a configuration~$\gamma' \in \Gamma(S)$ via problem-solving or learning that allows it to perform task~$T$. If no suitable configuration exists or cannot be found to perform task $T$, then we set
$$
 c^{\mathrm{ada}}(S,\gamma,T) = \infty.
$$
\end{definition}

If the task-performing system we are interested in is in fact the learning system, then we will often find that reconfiguration cost are negligible compared to learning cost. Adaption cost then effectively reduce to learning cost. Hence, we can also consider learning cost as a special case of adaption cost. 

\begin{figure}
    \centering
\begin{tikzpicture}[
    scale=5,
    axis/.style={very thick, ->, >=stealth'},
    important line/.style={thick},
    dashed line/.style={dashed, thin},
    pile/.style={thick, ->, >=stealth', shorten <=2pt, shorten
    >=2pt},
    every node/.style={color=black}
    ]
    % axis
    \draw[axis] (-0.1,0)  -- (1.1,0);
    \node(xline)[below] at (0.5,0) {Configuration parameter $1$};
    \draw[axis] (0,-0.1) -- (0,1.1);
    \node(yline)[above, rotate=90] at (0,0.5) {Configuration parameter $2$};
    % Lines
    \draw[important line] (.45,.45) coordinate (A) -- (.65,.85)
        coordinate (B);
    \draw[important line] (.85,.45) coordinate (C) -- (.65,.85)
        coordinate (D);
    \draw[important line] (.45,.45) coordinate (E) -- (.85,.45)
        coordinate (F);
    % Intersection of lines
    \fill[black] (intersection cs:
       first line={(A) -- (B)},
       second line={(C) -- (D)}) coordinate (G) circle (.4pt)
       node[above,] {$T_2$};
    \fill[black] (intersection cs:
       first line={(A) -- (B)},
       second line={(E) -- (F)}) coordinate (H) circle (.4pt)
       node[left,] {$T_1$};
    \fill[black] (intersection cs:
       first line={(C) -- (D)},
       second line={(E) -- (F)}) coordinate (I) circle (.4pt)
       node[right,] {$T_3$};
    \path (H) edge[pil, dashed, bend right=30] (G);
    % The E point is placed more or less randomly
    \fill[black]  (E) +(-.1cm,-.3cm) coordinate (out) circle (.4pt);
    \path (out) edge[pil, dashed, bend right=30] (G);
    %    node[below left] {$B$};
    % Line connecting out and ext balances
    %\draw [pile] (out) -- (intersection of A--B and out--[shift={(0:1pt)}]out)
    %    coordinate (extbal);
    %\fill[red] (extbal) circle (.4pt) node[above] {$T_2$};
    % line connecting  out and int balances
    %\draw [pile] (out) -- (intersection of C--D and out--[shift={(0:1pt)}]out)
    %    coordinate (intbal);
    %\fill[red] (intbal) circle (.4pt) node[above] {$T_3$};
    % line between out og all balanced out :)
    %\draw[pile] (out) -- (E);
\end{tikzpicture}
    \caption{System adaption starting from some arbitrary system configuration compared to system adaption from a configuration that is optimal for a previous task $T_1$.}
    \label{fig:system_adaption}
\end{figure}
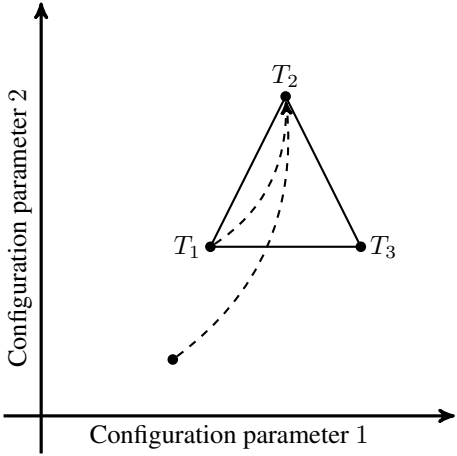

When a system adapts, the crucial difference to classical learning or problem-solving is how one gets to the configuration $\gamma$ in Definition~\ref{defi:learning_cost} from which the learning system starts the search in the configuration space $\Gamma(S)$. If it was arbitrarily fixed or chosen at random, then we would in fact be considering a classical learning or problem-solving situation. However, with adaption, we are interested in situations where the configuration~$\gamma$ results from the fact that system $S$ has already been configured for another task~$T_1$, or more generally, that the system has already been performing a sequence of tasks
$$
 H_{n} = (T_1,\dots,T_n),
$$
from the context~$W$ which we call a \emph{task history of length $n$}. There are several possibilities how a task history may arise. For instance, it may be an arbitrary but fixed selection of tasks from the context. The tasks in the history may also be drawn independently at random from the context, which makes $H_n$ a random vector. A random task history may also be constructed from a Markov decision process on the task set $\mathcal T$ or a subset thereof such that a task $T_j$ follows a task $T_i$ with some transition probaility.

A task history gives the learning system a chance to gain experience and knowledge about the context~$W$ which is then reflected in the system configuration $\gamma$. For instance, the system configuration $\gamma$ might already be in the right region of the configuration space or the system might have developed a better search strategy to find a suitable configuration for the new task $T$, 
see Figure~\ref{fig:system_adaption} for an illustration. However, if the new task~$T$ is completely unrelated to the tasks in the task history, then using the knowledge gained from the task history can also worsen the search compared to a random search. In any case, we expect that the adaption cost depend on the task history. This leads us to the following definition.   

\begin{definition}\label{defi:adaption_cost}
Let $n \in \mathbb{N}$. Given a task history~$H_n$ of length~$n$ and a task~$T$, the \textbf{adaption cost}
$$
 c^{\mathrm{ada}}(S, H_{n}, T)
$$
are the search and reconfiguration cost that system $S$ generates to be able to perform task~$T$ if it has previously been able to perform the tasks in the task history~$H_n$. For $n=0$, we let~$H_0 = \{\}$ be the empty set and define $c^{\mathrm{ada}}(S, H_{0}, T)$ to be the search and reconfiguration cost that system $S$ generates to learn to perform the task~$T$ from scratch, i.e., from the initial configuration~$\gamma_0$.
\end{definition}

Equipped with a formal definition of adaption cost, we can now define adaptability in a precise way.

\begin{definition}
\label{defi:adaptability}
Let $n \in \mathbb{N}$. Given a random task history~$H_n$ of length $n$ and a random task~$T$ from the context~$W$, we define the \textbf{adaptability} $\alpha_{W,n}(S)$ of system~$S$ with a task history of length $n$ from the context~$W$ as
$$
  \alpha_{W,n}(S) = -\mathbb{E}[c(S, H_{n}, T)].
$$
\end{definition}

Note that $\alpha_{W,0}(S)$ quantifies the ability of the task-performing system~$S$ to learn to perform a random task from the task context~$W$ \emph{from scratch}. Since the task history is random in Definition~\ref{defi:adaptability}, one would in general expect that the adaptability increases with the size of the task history. Ultimately, if the learning system is powerful enough and perfectly capable to learn the structure of the task context~$W$, then we expect that the adaptability converges to the average-case reconfigurability,
$$
 \lim_{n \to \infty} \alpha_{W,n}(S)  = \rho_W^{\mathrm{avg}}(S),
$$
as the learning cost decrease to zero. The reconfigurability is thus an upper bound for the adaptability (and the reconfiguration cost a lower bound for the adaption cost).

\subsubsection{Task diversity}

Adaptability and reconfigurability are essentially about the process of switching from one task to another. We may certainly expect that the more similar tasks are, the easier it is to switch between them. Likewise, we can expect higher adaption cost for switching between more dissimilar tasks. However, we cannot read off from the flexibility measures that we have defined in Definition~\ref{defi:reconfigurability} and Definition~\ref{defi:adaptability} how good a task-performing is in exploiting similarities between tasks and how well it can cope with dissimilar tasks. This section is thus devoted to developing measures of \emph{task diversity} that capture and quantify the second aspect of the informal flexibility notion given in Definition~\ref{defi:flexibility}.

Let $S$ be a task-performing system with some initial configuration $\gamma_0 \in \Gamma(S)$. Consider the situation where the system sequentially has to perform tasks $T_1,\dots,T_n$ from a given tasks context~$W$. Optimizing the system to perform task $T_1$ puts it into a configuration $\gamma_1 \in \Gamma(S)$. This will have some initial configuration and search cost
$$
 c^{\mathrm{ada}}(S,H_0, T_1),
$$
where $H_0=\{\}$ is the empty task history. The system is then equipped with a task history $H_1 = (T_1)$ and performing the task has execution cost~$c^{\mathrm{run}}(S,\gamma_1)$. Optimizing the system for the subsequent task $T_2$ puts it into a system configuration~$\gamma_2 \in \Gamma(S)$ at adaption cost
$$
 c^{\mathrm{ada}}(S,H_1,T_2).
$$
Performing task $T_2$ has cost~$c^{\mathrm{run}}(S,\gamma_2)$
and its task history is now
$$
 H_2 = (T_1, T_2).
$$
Iteratively proceeding in this way until the system has performed all tasks $T_1,\dots, T_n$ generates the following total cost.

\begin{definition}
\label{defi:total_cost}
Let $S$ be a task-performing sytem with initial configuration $\gamma_0 \in \Gamma(S)$.
Let $T_1,\dots, T_n$ be a sequence of tasks from the context~$W$. Let
$$
 H_0 = \{\}
$$
and, for $i=1,\dots,n$, let
$$
 H_i= (T_1,\dots,T_i).
$$
Furthermore, for $i=1,\dots,n$, let $\gamma_i \in \Gamma(S)$ be the configuration resulting from adapting the system to task $T_i$.
The \textbf{total cost} of performing the task sequence $H_n$ is given by
$$
 c^{\mathrm{tot}}(S,H_n) := \sum_{i=1}^n \left(c^{\mathrm{ada}}(S,H_{i-1},T_i) + c^{\mathrm{run}}(S,\gamma_i)\right).
$$
\end{definition}

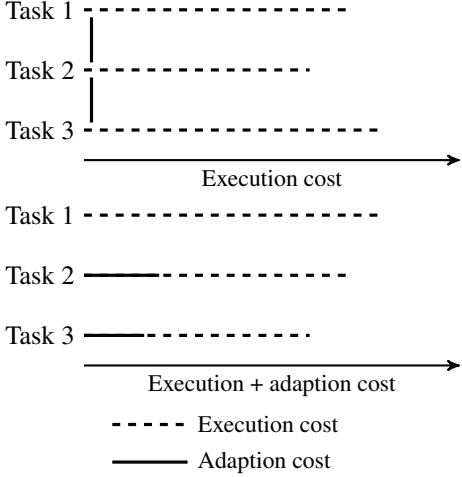
\begin{figure}
    \centering
    \begin{tikzpicture}[scale=1.0]
    % Draw axes
    \draw [->, thick] (0,0) -- (5,0);
    \node(xaxis)[below] at (2.5,0) {\small Execution cost};

    \draw[very thick,dashed] (0,2) -- (3.5,2);
    \draw node [left] at (0,2) {Task 1};
    
    \draw[very thick] (.1,1.9) -- (.1,1.3);
    
    \draw[very thick, dashed] (0,1.2) -- (3,1.2);
    \draw node [left] at (0,1.2) {Task 2};
    
    \draw[very thick] (.1,1.1) -- (.1,.5);
    
    \draw[very thick,dashed] (0,.4) -- (4,.4);
    \draw node [left] at (0,.4) {Task 3};
    \end{tikzpicture}
    
    \begin{tikzpicture}[scale=1.0]
    % Draw axes
    \draw [->, thick] (0,0) -- (5,0);
    \node(xaxis)[below] at (2.5,0) {\small Execution + adaption cost};

    \draw[very thick, dashed] (0,2) -- (4,2);
    \draw node [left] at (0,2) {Task 1};
    
    \draw[very thick, dashed] (0,1.2) -- (3.5,1.2);
    \draw[very thick] (0,1.2) -- (1,1.2);
    \draw node [left] at (0,1.2) {Task 2};
    
    \draw[very thick, dashed] (0,.4) -- (3,.4);
    \draw[very thick] (0,.4) -- (.8,.4);
    \draw node [left] at (0,.4) {Task 3};
    \end{tikzpicture}
    
    \begin{tikzpicture}
        \draw[very thick, dashed] (0,0.5) -- (1,0.5) node[right] {\small Execution cost};
        \draw[very thick] (0,0) -- (1,0) node[right] {\small Adaption cost};
    \end{tikzpicture}
    \caption{Necessary system adaptions (top) vs. system specialization on-the-fly (bottom).}
    \label{fig:total_cost}
\end{figure}

Given a fixed total cost budget, we are interested in system designs for which the number of tasks that can be performed with that budget is as large as possible. Depending on the scenario under consideration, the role of the adaption to drive up the number of tasks can be different. If each task necessarily requires a different system configuration, then a system adaption is strictly necessary. In this case, minimizing the adaption cost is an endeavour that must be made irrespective of any needs to minimize the execution cost. This scenario frequently occurs in manufacturing. However, there are also scenarios where the same system configuration, say $\bar \gamma \in \Gamma(S)$, works for all tasks but in an suboptimal way. Then, system adaption is not strictly necessary but can be still be used to achieve a \emph{system specialization on-the-fly}. The goal of the adaption is then to find a configuration $\gamma_i$ such that
$$
 c^\mathrm{ada}(S,H_{i-1}, T_i) + c^{\mathrm{run}}(S,\gamma_i) < c^{\mathrm{run}}(S,\bar \gamma).
$$
The minimization of adaption cost is in this case part of the minimization of execution cost.
A practical example for this scenario is instance-based algorithm selection for NP-hard optimization problems~\cite{kerschke2019automated}.

In biology, the simplest diversity measure is species richness, which simply is the number of different species in a population. Following this definition, we define the task richness of a task-performing system in a given context as the number of \emph{representative} tasks from the context that the system can perform with a given total cost budget.

\begin{definition}
\label{defi:task_richness}
For $n \in \mathbb{N}$, let $H_n =(T_1,\dots,T_n)$ be a sequence of independent random tasks from a given context~$W=(\mathcal T,Q)$ and let $b >0$ be a total cost budget. Then, we define
the \textbf{task richness} of task-performing system $S$ for a given budget $b$ to be
$$
 \Delta_{W,b}(S) =  \mathbb{E}\left[\max\{ n \in \mathbb{N}:  c^{\mathrm{tot}}(S,H_n) < b \}\right].
$$
\end{definition}

With task richness, we implicitly assume that all tasks are equally dissimilar. If we have a distance measure $d$ on the space of tasks $\mathcal T$ at hand that allows for a quantification of task relatedness, then we can generalize Definition~\ref{defi:task_richness} as follows.

\begin{definition}
\label{defi:task_diversity}
For $n \in \mathbb{N}$, let $H_n =(T_1,\dots,T_n)$ be a sequence of independent random tasks from a given context~$W=(\mathcal T, Q)$ and let $b >0$ be a total cost budget. Furthermore, let
$$
 \ell(H_n) = 1 + \sum_{i=1}^{n-1} d(T_i,T_{i+1}).
$$
Then, we define the \textbf{task diversity} of system $S$ for a given budget $b$ to be
$$
 \Delta_{W,d,b}(S) =  \mathbb{E}\left[\max\{ \ell(H_n) :  c^{\mathrm{tot}}(S,H_n) < b \}\right].
$$
\end{definition}

\begin{figure}
    \centering
    \begin{tikzpicture}[scale=1.5]
    % Draw axes
    \draw [->, thick] (0,0) -- (0,2.5);
    \node(yline)[above, rotate=90] at (0,1.2) {\small Task diversity $\Delta_{W,d,b}(S)$};
    \draw [->, thick] (0,0) -- (3,0);
    \node(xaxis)[below] at (1.5,0) {\small Total cost budget $b$};
    
    \draw (0,0) -- (2.2,1) node [right] {System $S_1$};
    \draw (0,0) -- (2.2,1.5) node [right] {System $S_2$};
\end{tikzpicture}
    \caption{System designs where task diversity grows faster when increasing the total cost budget are better optimized for a given context.}
    \label{fig:task_diversity}
\end{figure}
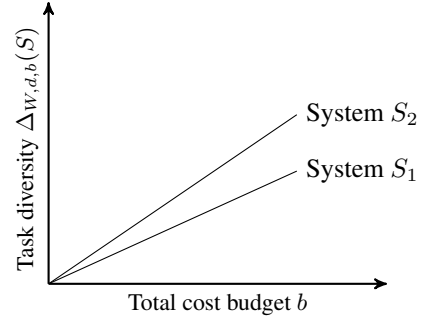

Based on this definition of task diversity, we can say that one system design is more suitable or better optimized for a context than another system design if the task diversity of the former grows faster with the available total cost budget, see Figure~\ref{fig:task_diversity} for an illustration.

Note that task relatedness is inherently difficult to measure. In the end, the relatedness of tasks is given by the similarity of the algorithms and processes that solve the tasks. Quantifying task relatedness rigorously in advance is thus hardly possible. Proxy measures can be derived from human experts evaluating task relatedness but these measures are not guaranteed to be correct. Developing formal notions of task relatedness is an ongoing research topic in statistical learning theory~\cite{tripuraneni2020theory}.

$$
%while
%$$
% \mathbb{E}[c(S,H_n^{\mathrm{MVG}}, T)] = 3.
$$
%Moreover, the reconfiguration cost in the MVG scenario are on average five times smaller than in the FG scenario. The reason for the better adaptability and reconfigurability in the MVG scenario is that the switching between the tasks in the history $H_n^{\mathrm{MVG}}$ lets the genetic algorithm search in a region of the configuration space $\Gamma(S)$ where the corresponding wirings between the NAND gates reflect the hierarchical and modular structure of goals, which is not the case in the FG scenario.  

%\subsection{Learning to learn.}

%\todo{tidy up}

%One way to understand learning to learn from the viewpoint of system flexibility:
%\begin{itemize}
%    \item System complexity is given by the choice of machine learning model and its number of parameters.
%    \item In the beginning, the system configuration is typically random.
%    \item Adaption cost is given by the amount of training data needed for new tasks
%    \item Then learning to learn means to increase task diversity and system adaptability over time.
%\end{itemize}

\subsection{Related aspects of system design and optimization}
\label{sec:related}

In this section, we discuss relations of system flexibility to other aspects of system design and optimization.

\subsubsection{Robustness}

A system is robust if it continues to function in the face of perturbations~\cite{wagner2007robustness}. Depending on the context and the system under consideration, the nature of a perturbation can be quite different and it can be considered small or large. Typical small perturbations are erroneous or noisy inputs or errors occurring during process execution. Typical examples for large perturbations are damages or failures within system components or drastic changes in requirements or environmental conditions. For small perturbations, system robustness is closely related to system stability~\cite{hirsch2012differential}. Depending on the field, robustness to large perturbations is also referred to as resilience~\cite{holling1973resilience,laboy2016resilience} or fault tolerance~\cite{isermann2005fault}.

In general, we can distinguish two different approaches to robustness. The first seeks for system designs that allow the system to cope with perturbations in a static fashion without changing its configuration.  This corresponds more to the daily use of the term `robust' and in some areas of engineering, `robust design` solely refers to this first approach, e.g., in controller design~\cite{zhou1998essentials,lin2007robust} or supply chain risk management~\cite{wieland2012dealing}. In the second approach, a system achieves robustness by adapting its configuration to the changed situation caused by a perturbation. Here we have an obvious connection to system flexibility, in particular, in the case of large perturbations. Sustaining a certain function is the overarching goal and perturbations lead to variations in the tasks that have to be performed to reach the overarching goal. System flexibility on a lower level can thus be a means to make a system robust on a higher level. High adaptability means that the system can quickly restore its function in the face of perturbations, and high task diversity supports the system in coping with a large variety of perturbations.

Concerning the search in a configuration space, system flexibility and robustness on the same level can be in conflict. While it is favorable for robustness that many configurations lead to the same system functionality, it is more difficult to find novel system functions under such circumstances. The monograph~\cite{wagner2007robustness} discusses how the structure of search spaces can mitigate this conflict in the evolution of biological systems such that evolvability and robustness can go hand in hand, see also Section~\ref{sec:degeneracy}.

\subsubsection{Multi-objective optimization}

Assume to be given~$n$ tasks~$T_1,\dots,T_n$ with associated performance measures~$P_1,\dots,P_n$ (this includes the case where all tasks are basically the same expect that we want to optimize different performance criteria or objectives). Leaving aside any feasibility constraints, the multi-objective optimization problem of finding a system configuration that is optimal for all~$n$ tasks simultaneously can mathematically be formulated as
$$
\max_{\gamma \in \Gamma(S)} (P_1(\gamma),\dots, P_n(\gamma)),
$$
where the maximum is taken component-wise and
$$P_i(\gamma) := P_i(S(\gamma))$$
is the performance of system~$S$ in doing tasks~$T_i$ when it is in configuration~$\gamma$.

Typically, there will be no configuration that maximizes all performance measures simultaneously. Therefore, one is interested in optimal \emph{trade-offs} between the tasks performances. Concretely, one is interested in \emph{Pareto-optimal} configurations, which are configurations that cannot be improved for any of the tasks without degrading the performance in at least one of the other tasks. In mathematical terms, a system configuration $\gamma \in \Gamma(S)$ is said to \emph{Pareto-dominate} another system configuration $\gamma' \in \Gamma(S)$ if
$$
 P_i(\gamma') \leq P_i(\gamma) 
$$
for all~$i \in \{1,\dots,n\}$ and
$$
P_j(\gamma') < P_j(\gamma)
$$
for at least one $j \in \{1,\dots,n\}$. Now a system configuration is called Pareto-optimal if there does not exists another configuration that dominates it. The set of all Pareto-optimal configurations is called the \emph{Pareto front}. If no preferences with regard to the tasks are available, then solving the multi-objective optimization problem is understood as approximating or computing all or a representative set of the the Pareto front~\cite{ehrgott2005multicriteria}.

It is common to scalarize the multi-objective optimization problem~\cite{jahn1985scalarization}. For scalarization, one chooses a scalar fitness function
$$h(\gamma) = h(P_1(\gamma),\dots,P_n(\gamma))$$
such that configurations maximizing $h$ are Pareto optimal for the original multi-objective problem. The fitness function $h$ can be used to incorporate importance differences with regard to the tasks. Under certain assumptions on the fitness function~$h$ and the performance measures~$P_1,\dots,P_n$, one can show that the Pareto front takes the form of a polytope with~$n$ vertices in the configuration space $\Gamma(S)$, where vertex~$i$ is the configuration that maximizes~$P_i$~\cite{shoval2012evolutionary}.

Let us now discuss the difference in focus between system flexibility and multi-objective optimization based on the example in Figure~\ref{fig:system_adaption} where we considered three tasks. Under the assumptions described in the previous paragraph, the Pareto front forms a triangle. With multi-objective optimization, our interest is the triangle as such. For instance, we could be interested to determine its position in the configuration space. Given certain preferences with regard to the tasks, we could also be interested in finding the optimal trade-off configuration in the triangle reflecting the preferences. Ideally, we would like to have a system design that makes the triangle as small as possible such that we have to make less severe trade-offs. In contrast, with system flexibility, we are interested in system designs that allow us to travel as cheaply as possible within the triangle. In a mixed scenario of multi-objective optimization and flexibility, the interest could be to have means to recalculate the Pareto front efficiently when conditions change that affect all considered tasks.

\section{The elements of flexibility}
\label{sec:elements}

Solving a flexibility optimization problem such as generally described by Definition~\ref{defi:flexibility_optimization} involves a search in a space of task-performing systems. A search step will in many cases be very costly as it means to evaluate the performance of a task-performing system on many different tasks or task variations and the evaluation on one single task can already be costly, in particular for real-world tasks. Hence, to avoid costly evaluations as much as possible, we are interested to restrict the search to system designs that generally have a good chance to yield high system flexibility.

The way in which living systems have evolved suggests that such system designs exist, at least for the tasks and task contexts living systems are exposed to by natural environments. As we have already described in Section~\ref{sec:evolvability}, there is a strong overlap between the structural and behavioral characteristics of biological systems responsible for evolvability, phenotypic plasticity, and the flexibility of human learning. It seems that no matter what level of biological organization and what kind of tasks we consider, we always find the following aspects in the behavior and design of biological systems that promote flexibility:

\begin{enumerate}
    \item Hierarchy,
    \item Modularity,
    \item Weak regulatory linkage,
    \item Exploration,
    \item Degeneracy and neutrality,
    \item Weak links.
\end{enumerate}

We call these the \emph{elements of flexibility}. In the following, we properly introduce these elements and give an overview of how they appear in biological systems and promote flexibility. This is intended primarily as a source of inspiration and a motivation to think in a bio-inspired way about the potential to incorporate the elements of flexibility in man-made task-performing systems. As any form of adaption in a man-made task-performing systems involves some form of learning and adjusting of model structures and/or parameters, it is a debatable point if a distinction between evolvability, phenotypic plasticity, and learning flexibility makes sense for man-made task-performing systems. Hence, to keep the presentation of the elements of flexibility concise, we focus mainly on evolutionary aspects in the following.

\subsection{Hierarchy}
\label{sec:hierarchy}

A \emph{hierarchical system} can be defined as system that is composed of interrelated subsystems, each of which is hierarchical again until some lowest level of subsystem is reached~\cite{simon1962architecture}. A special form of hierarchy arises from \emph{recursion}. Here the same subsystems appear on each level of the hierarchy. Likewise to systems, processes and tasks can be hierarchical. A process being hierarchical means that there is an algorithmic description of the process that has a hierarchical structure. Formal definitions of hierarchy can for example be given based on graph theory~\cite{corominas2011hierarchy}. A \emph{hierarchical decomposition} decomposes a given system, process or task into a hierarchy of interrelated elements. This yields a hierarchical view or representation of the system, process or task, which does not have to be unique. A hierarchical decomposition usually leads to a description of increasing detail along the levels of the hierarchy. In a \emph{hierarchical composition}, systems are built by combining primitive elements which in turn can be combined to create more complex systems, and so on.

%Hierarchical organization is ubiquitous in biological systems. Examples include brain networks~\cite{zhou2006hierarchy,meunier2010modular}, metabolic networks~\cite{ravasz2002metabolic}, gene regulatory networks~\cite{yu2006regulatory}, and the hierarchical refinement of the body plan during embryogenesis~\cite{kirschner2006plausibility,carroll2013dna}. We discuss the later in more detail below.

%There are several hypotheses why hierarchy is omnipresent in biological systems and why they have evolved in such a way.
%\begin{itemize}
%    \item Hierarchy allows for stable intermediate states and evolution times logarithmic in the system complexity~\cite{simon1962architecture}.
%    \item Hierarchy allows a reduction of environmental uncertainty through the construction of dynamical processes with a range of characteristic time constants to separate between slowly and quickly adapting parts~\cite{flack2013timescales}.
%    \item Hierarchy may arise as a byproduct of random processes~\cite{corominas2013origins}.
%    \item Hierarchy is a result of resource constraints, e.g., through connection costs~\cite{mengistu2016evolutionary}.
%\end{itemize}

%Most importantly from our point of view, hierarchy is thought to confer greater robustness and adaptability~\cite{meunier2010modular,meunier2009hierarchical,bassett2010efficient,mengistu2016evolutionary}.

Hierarchy has long been identified as an important feature of complex biological systems, contributing both to their robustness and adaptability \cite{simon1962architecture,bassett2010efficient,meunier2010modular}. For example, Simon \cite{simon1962architecture} used the watchmaker parable already in 1962 to argue that it is much easier and faster to evolve hierarchical systems, by first evolving the elementary subsystems and then building up on and reusing them when necessary. Some examples of the presence of hierarchy in complex biological systems are given below:

\begin{itemize}

    \item In the metabolic network of E. coli, that model reactions between substrates inside a cell, the properties of scale-free distributions (only a few nodes with many connections) and modularity (high clustering coefficient) coexist in a hierarchical structure formed by highly interconnected modules that combine in larger modules, both structurally and functionally \cite{ravasz2002metabolic}.

    \item The gene regulatory networks from Escherichia coli and Saccharomyces cerevisiae, which model the interactions between transcription factors in a cell, and the network formed by the administrative organization of the governement of Macao, have a hierarchical structure where the mid-level nodes are more essential \cite{yu2006regulatory}. Regulatory networks are also part of a higher hierarchy, as especially the top-level nodes receive signals from external agents.

    \item Brain networks were shown to present the property of hierarchical modularity both in the anatomical as well as function levels (cortical network of the cat brain, where a node is a cortical area and edges are interactions between cortical areas \cite{zhou2006hierarchy}, and human brain functional networks, where nodes correspond to brain regions and connections are made between nodes if their functional time series are sufficiently correlated \cite{meunier2009hierarchical}).

\end{itemize}

\begin{figure}
    \centering
    \includegraphics[width=5cm]{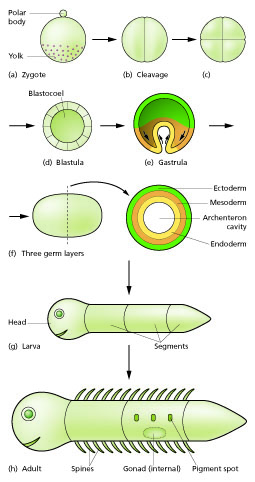}
    \caption{Selected stages of the hierarchical refinement of the body plan during embryogenesis.\protect\footnotemark}
    \label{fig:hierarchical_refinement}
\end{figure}

A sophisticated example for the occurrence of hierarchy in nature is the hierarchical refinement of the body plan during embryogenesis~\cite{kirschner2006plausibility,carroll2013dna}, see Figure~\ref{fig:hierarchical_refinement}. In the very early stages, relatively little spatial organization of the cells is given. This quickly changes and a spatial division of the embryo starts to occur. Cells start to migrate to different parts in order to form layers. These layers are divided into further segments such that a two-dimensional coordinate system is established in the embryo, which describes the high-level body axes. The segments in this coordinate system are the starting point for the development of secondary fields, in which precursors and initial structures for organs and appendages begin to grow such as the heart, the nervous system and limbs. With the secondary fields, additional, more fine grained coordinate systems are established to control positioning, number and identity of forming structures. This is followed by further divisions into subdomains and compartments.

\footnotetext{By JMWSlack - Own work, CC BY-SA 4.0, \url{https://commons.wikimedia.org/w/index.php?curid=40874860}}

A small group of genes, which is called the genetic toolkit of development, controls the developmental process. The formation of the first high-level body axis is controlled by five groups of genes of the toolkit that influence different parts of the axis formation. Other groups of genes control the formation of the second high-level body axis. A very important role for the secondary fields play the Hox genes. This is a family of genes which act as selector genes to initiate the development of certain organs, cell differentiations or compartments. It is important to understand that these genes themselves do not describe how to develop certain body parts but only trigger developmental modules located elsewhere in the genome. An often-observed phenomenon is that the order of the selector genes in the genome corresponds to the order of the segments that they influence, which is called collinearity~\cite{carroll2013dna}.

The genetic toolkit of development is highly conserved across all animal species. Some genes, though not exactly the same, can even be interchanged between animal species as different as a fly and a mouse~\cite{carroll2013dna}. This suggests that evolution once has found a program superior to all other possible ways of how to develop organisms. This program is since then only varied, refined, and enhanced. It is an elaborate combination of hierarchy, modularity (see Section~\ref{sec:modularity}), and exploratory behavior (see Section~\ref{sec:exploration}) that helps organisms to adapt during evolution~\cite{kirschner1998evolvability} but also to adjust to environmental conditions within their lifespan~\cite{whitman2009phenotypic}.

The origins of hierarchy during evolution is still an ongoing reasearch topic. The work in \cite{corominas2013origins} defines a three dimensional space of directed networks in order to characterize how hierarchical a network is. Both randomly generated artificial networks and a number of real networks from biology, society, and technical systems have a bow-tie architecutre, are not ordered, and are not feedforward, lying in the same region of this space. This suggests that hierarchy in nature can be in many cases a byproduct of randomness and of how networks are generated instead of a result of evolutionary advantages or selection pressure. Some networks, however, lie outside this region, as gene regulatory networks, ecological networks, and electronic circuits, suggesting that some form of selection pressure may be at work.

Flack et al. \cite{flack2013timescales} argue that higher levels in the hierarchy of a system can act as slow variables, which are variables that serve as aggreate or average measures over a larger number of lower level system variables. As such, slow variables tend to oscilate less over time, and a larger change in lower level variables is required for that to result in a change in a higher level, or slow, variable. On one side, slow variables reduce uncertainty and contribute to the robustness of a system – as they are more stable, selection can act on them instead of tracking every change on lower level variables. On the other hand, they also provide feedback to and control over lower level variables, facilitating variation, making it possible to selection act on different levels, and thus contributing to evolvability.

In \cite{mengistu2016evolutionary}, Mengistu et al. suggest that a selection pressure towards minimizing connection costs in networks is one of the explanations for the emergence of hierarchy in biological systems. A selection pressure on connection cost is biologically plausible, given that some networks, like the brain, are optimized in regards to connection cost. They conduct experiments where an evolutionary algorithm evolves a neural network to match the input and output of a simple boolean circuit. When optimizing both for quality and lower connection cost, the circuits evolved present higher modularity and hierarchy and converge faster. There is also an increase in functional hierarchy, where individual nodes in the solutions are responsible for solving subtasks. Furthermore, the obtained modular and hierarchical solutions also adapt faster to solving a related task (a circuit with a related pattern). All of these observations also hold when the system is configured to produce solutions that are hierarchical but not modular, showing that the effects of hierarchy are additive to modularity and can be isolated.

\subsection{Modularity}
\label{sec:modularity}

Modularity is studied in many areas with differing definitions that have a common gist, however. In the strongest sense, we call a subsystem a \emph{module} if it has the following properties:

\begin{enumerate}
    \item \emph{Separation of function}: each module has a distinct function and thus carries out a separable subtask within the system.
    \item \emph{Reusability}: The subtask solved by a module appears frequently in the overall task or in many different tasks.
    \item \emph{Recombinability}: The module can be usefully combined with other modules to form new systems capable of solving different tasks.
    \item \emph{Encapsulation of detail}: Communication with the part happens via an interface such that internal details can be unknown to other modules. There is typically more intra-module interaction than inter-module interaction.
\end{enumerate}

\emph{Modularization} is the process to decompose a given system into modules.

\begin{figure}
    \centering
    \begin{tikzpicture}[node distance=1.5cm]
    \tikzstyle{every state}=[fill=white,draw=black,text=black]
        \node[state] (a_x) at (0.75,5) {$x$};
        \path (a_x) edge[pil,loop above] (a_x);
        \node (a_label) at (0.75,3.5) {(a)};
        
        \node[state] (b_x) at (4,6) {$x$};
        \node[state] (b_z1) [below left of=b_x] {$z_1$};
        \node[state] (b_z2) [below of=b_x] {$z_2$};
        \node[state] (b_z3) [below right of=b_x] {$z_3$};
        \path (b_x) edge[pil, loop above] (b_x)
              (b_x) edge[pil] (b_z1)
              (b_x) edge[pil] (b_z2)
              (b_x) edge[pil] (b_z3);
        \node (b_label) at (4,3.5) {(b)};
        
        \node[state] (c_x1) at (0,2) {$x_1$};
        \node[state] (c_x2) [right of=c_x1] {$x_2$};
        \node[state] (c_z1) [below of=c_x1] {$z_1$};
        \node[state] (c_z2) [below of=c_x2] {$z_2$};
        \path (c_x1) edge[pil] (c_z1)
              (c_x1) edge[pil] (c_z2)
              (c_x2) edge[pil] (c_z1)
              (c_x2) edge[pil] (c_z2);
        \node (c_label) at (0.75, -1.5) {(c)};
        
        \node[state] (d_x) at (4,2) {$x$};
        \node[state] (d_y) [below of=d_x] {$y$};
        \node[state] (d_z1) [below left of=d_y] {$z_1$};
        \node[state] (d_z2) [below right of=d_y] {$z_2$};
        \path (d_x) edge[pil] (d_y)
              (d_x) edge[pil] (d_z1)
              (d_x) edge[pil] (d_z2)
              (d_y) edge[pil] (d_z1)
              (d_y) edge[pil] (d_z2);
        \node (d_label) at (4,-1.5) {(d)};
        
        \node[state] (e_x1) at (0,-3) {$x_1$};
        \node[state] (e_x2) [right of=e_x1] {$x_2$};
        \node (e_xdots) [right of=e_x2] {...};
        \node[state] (e_xn) [right of=e_xdots] {$x_n$};
        \node[state] (e_z1) [below of=e_x1] {$z_1$};
        \node[state] (e_z2) [below of=e_x2] {$z_2$};
        \node (e_zdots) [below of=e_xdots] {...};
        \node[state] (e_zn) [below of=e_xn] {$z_n$};
        \path (e_x1) edge[pil] (e_z1)
              (e_x1) edge[pil] (e_z2)
              (e_x2) edge[pil] (e_z1)
              (e_x2) edge[pil] (e_zn)
              (e_xn) edge[pil] (e_z1)
              (e_xn) edge[pil] (e_z2)
              (e_xn) edge[pil] (e_zn);
       \node (e_label) at (2.375,-5.5) {(e)};
    \end{tikzpicture}
    \caption{Examples for the four basic functional module types of sensory transcription networks in cells. (a) autoregulation, (b) single input module (SIM), (c) bi-fan, (d) two-output feedforward loop, (e) dense overlapping regulons (DOR).}
    \label{fig:network_motifs}
\end{figure}
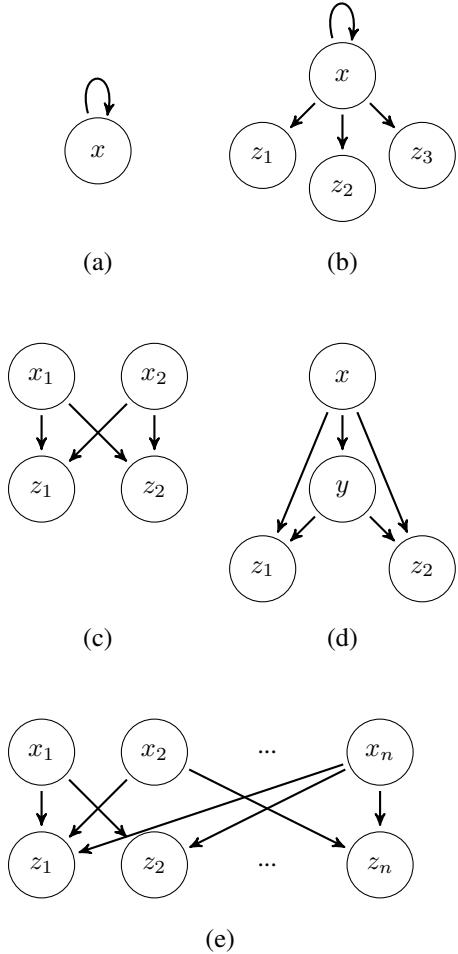

Modularity is a well known property of complex biological systems~\cite{melo2016modularity}, which can mostly be decomposed into subsystems where inter-system interactions are less frequent than intra-system interactions~\cite{simon1962architecture}. In turn, changes in specific subsystems have a lower chance of negatively affecting the whole organism, promoting robustness and evolvability~\cite{kirschner1998evolvability}. Some examples of modularity in biological systems are:

\begin{itemize}
    \item Different cell types have specialized functions in the organism, and form different organs which have specific functions~\cite{wagner2007road}.
    \item Neuronal networks in animal brains have modular structures - some motifs appear more frequently than others~\cite{sporns2004motifs}.
    \item If one models protein-protein interactions as a "protein network", in which nodes correspond to proteins and edges to interactions, it is possible to identify modules - sets of highly connected proteins with less connections to other sets~\cite{wagner2007road}.
    \item Some evolutionary traits can also be viewed as modules, as, for example, some beak traits in birds, which were evolved independently from each other~\cite{wagner2007road}.
    \item The body plan of all metazoans is similar in an early stage (phylotypic stage), composed of modules (conserved core processes) that can evolve to very different bodies~\cite{kirschner1998evolvability}, see also Section~\ref{sec:hierarchy}.
    \item Sensory transcription networks in cells are basically build from only four types of functional modules~\cite{alon2019introduction}, see Figure~\ref{fig:network_motifs}.
\end{itemize}

In contrast to technical systems, identifying modules in biological systems is a reverse engineering process that requires measures to quantify the presence or absence of the different aspects of modularity. Reusability can be derived from reoccurrence of structures in a biological system, e.g., via network motif identification~\cite{milo2002network,alon2007network}.  Encapsulation of details can be measured by clustering techniques or methods to find cliques or communities in graphs~\cite{hofman2008bayesian,ziv2005information}. Function identification is certainly the most challenging aspect that requires assessment by human experts or costly algorithmic approaches, e.g., subnet output comparison to reference values~\cite{velez2016identifying}. The debate what constitutes modules in biological systems is not settled~\cite{winther2001varieties}. While reusability, recombinability and encapsulation of detail is frequently encountered, separation of function does often not apply to modules in biological systems, see Section~\ref{sec:degeneracy}. From the examples given above one can differentiate three types of modularity at different levels of biological organization: variational modules are sets of features that vary together throughout evolution; functional modules are sets of features that together perform a physiological function; developmental modules are elements responsible for pattern formation and differentiation during development~\cite{wagner2007road,bolker2015modularity}. Traits that are varied together tend to be dedicated to the same function, which means that variational modules correspond often also to functional modules. 

Although there is abundant proof of the existence of modules in biological systems, it is still not clear how they arise during evolution. In ``neutral models'', modularity is not selected directly – for example, modularity can arise from gene duplication and differentiation~\cite{wagner2007road}. A scenario in which modularity is selected for is when it directly affects fitness during the development phase – modular systems present a developmental advantage. For example, evolving the topology of a neural network and then training the weights for the specific problem to be solved results in modular networks~\cite{wagner2007road}. Also, selection may favour robust systems, which selects modularity indirectly, as changes inside a module have less effect than changes between modules~\cite{wagner2007road}.

Interesting results from the viewpoint of system flexibility are the ones from \cite{parter2008faciliated,kasthan2005spontaneous}, which show that when systems are evolved under changing environments they tend to result in modular structures and are more flexible. For that, the authors use an evolutionary algorithm to evolve digital circuits and RNA structures under a regime where the objective function varies cyclically over modularly similar goals with the same structure but different modules along the evolution.

Solutions evolved that way can be easily adapted to solve other problems that were also used as objective during evolution, and solutions are often modular, with different modules being responsible for specific functions. Adapting from one solution to another requires only small changes to use one module instead of other, for example, and these changes consists mainly of mutating specific positions in the genotype. Additionally, mutations have a larger effect on the module to which they are applied but have small effects on other modules (reduced pleiotropy). Solutions, however, are not easily adapted to novel unseen goals, unless these novel goals consist of the same constrained structure of the ones used during evolution.

It is also possible for modularity to arise without changing environments~\cite{clune2013origins,chung2012structural,sporns2004motifs,melo2015directional}. In~\cite{clune2013origins} a neural network is evolved to detect objects on the left and right sides of a retina and compare between using only the performance or the performance plus a measure for the connection cost as fitness. Networks evolved for optimizing also the connection cost have both better performance and higher modularity. When the problem changes (from detecting if there is an object on the left OR right side to left AND right side, and vice-versa), the evolved modular networks also have a much greater evolvability, for example, needing 12 as opposed to 222 generations to adapt to the new task. In the context of network-based communications subsystem on an integrated circuit (network-on-chip) it has been confirmed that using higher weights for the cost factor in the fitness functions leads to more modular networks~\cite{chung2012structural}. 

The work~\cite{sporns2004motifs} obtains a result in the same direction for motifs in brain networks when studying structural motifs, which are specific nodes and edges patterns, and functional motifs, that are all motifs that can be derived given the constraints of a determined structural motif. When evolving networks using a selection pressure towards more functional motifs, the resulting networks are similar to brain networks in terms of number of structural motifs and which motifs appear with a higher frequency.

In conclusion, it seems that there are many mechanisms by which modules can arise and be selected for by evolution. As stated in~\cite{wagner2007road}, ``it seems that the origin of modularity requires both a mutational process that favours the origin of modularity and selection pressures that can take advantage of and reinforce the mutational bias''.

\subsection{Weak regulatory linkage}
\label{sec:wrl}

Weak regulatory linkage refers to design principles and mechanisms underlying the interface of a module or a system that facilitate its reconfiguration and recombination. Typical characteristics of interfaces obeying weak regulatory linkage are:

\begin{enumerate}[label=(\roman*)]
    \item Control signals are unspecific and can be imprecise.
    \item Control signals are non-instructive, only selecting between a number of predefined states of the regulated process.
    \item The integration of new signal types is possible without disturbing the existing input-output-relationship too much.
    \item Changes to existing control signals are possible without disturbing the existing input-output-relationship too much.
\end{enumerate}

Items (iii), (iv) are referred to as \emph{scalable integration} of many inputs into one, or many, outputs in~\cite{gunawardena2018wrt}.

In biological systems, linkage refers to the connections of processes to each other or to particular conditions~\cite{gerhart2007theory}. Regulatory linkage is the special case where one process controls the activity of another process, i.e., the output of the first process acts as a control signal to regulate the activity of the second process. \emph{Weak regulatory linkage} now refers to a number of evolved design principles and mechanisms for regulatory linkage in biological systems that allow the regulatory links to be `weak' such that the regulated processes can be easily reconfigured and recombined~\cite{conrad1990geometry,kirschner1998evolvability,gerhart2007theory,gunawardena2018wrt}. Typical characteristics of weak regulation are described by items (i)-(iv) above.

An example for an implementation of weak regulatory linkage is allostery~\cite{monod1965nature}. Allosteric proteins have a regulatory site, that binds to activators or inhibitors, and functional sites, that binds to a substrate to perform the protein's function. A variety of regulators bind to the regulatory site, either to activate the protein (activators) or to inhibit its activity (inhibitors). These regulators are independent from the protein's function, and they do not need to coevolve; different regulators can activate or inhibit the protein's activity, which reduces constraint and renders additional flexibility to the system. Allosteric proteins are omnipresent and have many functions in the organism, e.g., in metabolism, signal transduction pathways, and neuronal excitation.

Another frequently encountered realization of weak regulatory linkage are bow-tie architectures~\cite{csete2004bow,friedlander2015bowtie}. In a bow-tie, there is a large number of inputs called the fan-in, an unchanged mechanism called the knot that processes these inputs, and a large number of resulting outputs called the fan-out. The knot is able to process a large number of inputs and produce a large number of outputs, being robust to fluctuations in the inputs and versatile to be used in different situations to produce different outputs, by means of regulatory processes and feedback networks. An important example for a bow-tie architecture can be found in the metabolism. A large variety of nutrients can be synthesized into necessary building blocks for metabolism such as sugars, amino acids, nucleotides, and so on. Nutrients form the fan-in and the building blocks are the fan-out. Specific carriers and precursors metabolites act as the knot, the unchanged process that acts as starting point for the biosynthesis. Regulation and feedback make it possible for the same knot to act under different diets and to produce building blocks for specific purposes.

Bow-tie architectures can also be hiearchical. The amino acids and nucleotides produced as output in the metabolic bow-tie are used by a few universal polymerase molecules for the DNA translation and transduction processes. These polymerases act as the knot of the another bow-tie architecture, while genes and proteins act as the fan-in and fan-out, respectively. That is, the translation and transduction processes are robust to a variety of inputs and can produce specific outputs.

Weak regulatory linkage plays an important role for evolvability by reducing constrains between biological processes~\cite{kirschner1998evolvability}. Processes that interact are weakly linked do not need to coevolve; they do not need a specific chemical activator to perform its function, only something that initiates the process. Thus, activators can vary, change, and evolve without the need of simultaneously changing the process they activate, and the process itself can also evolve without the need of changing the activators at the same time. This makes systems more flexible and tolerant to variation, and increases evolvability by accommodating changes.

Computer experiments on the evolution of circuits and RNA secondary structures show that weak regulatory linkage can be the result of changing environmental conditions~\cite{parter2008faciliated}. The authors find that individuals evolved under a varying goal can be easily adapted to solve other problems that were also used as objective during evolution, and that these solutions are often modular, with different modules being responsible for specific functions. Adapting from one solution to another requires only small changes to use one module instead of other, for example, and these changes consists mainly of mutating specific positions in the genotype, which were called genetic triggers. This can be seen as a simple example of weak regulatory linkage, as it enables modules to be activated or deactivated. Furthermore, these modules have a conserved mechanism and an interface that allows them to interact with signals from different modules – they can be combined in different ways and evolved independently.

\subsection{Exploration}
\label{sec:exploration}

An \emph{exploratory process} is a process that generates a variety of outputs without a-priori information on the usefulness of the outputs and subsequently selects the most suitable one. The generation of outputs for exploration often involves a random process. A system that performs exploratory processes under certain conditions, e.g., in response to an environmental stimulus, exhibits \emph{exploratory behavior}.

Most fundamentally, natural evolution is an exploratory process. Genetic mutations generate a set of diverse organisms upon which natural selection acts. Environmental conditions and structural properties of organisms affect the efficiency of the evolutionary process, see also Section~\ref{sec:evolvability}. For example, digital circuits and RNA secondary structures evolved under a varying goal have a genetic and phenotypic neighbourhood with a high concentration of solutions to related problems that can be achieved by a small number of mutations~\cite{kasthan2005spontaneous,parter2008faciliated}. Degeneracy, where structurally different systems have the same or overlapping functions, contributes to the robustness and evolvability of organisms by means of neutral networks~\cite{whitacre2010degeneracy,draghi2010mutational}, see also Section~\ref{sec:degeneracy}.

Exploration also occurs within organisms. Specifically for certain cellular and developmental core processes, exploratory behavior has been defined as ``an adaptive behavior [...], wherein they generate many, if not an unlimited number of, specific outcome states, any of which can be stabilized selectively by other kinds of agents''~\cite{kirschner2006plausibility}. Examples for such exploratory processes are~\cite{kirschner1998evolvability,gerhart2007theory}:

\begin{itemize}
    \item The microtubules inside cells, that grow in random directions from a nucleation center. The ones that meet a stabilizing agent are trapped, while the others shrink back. Depending on where stabilizers are located, the spatial array of the microtubules will be different.
    \item The nervous system, where axons extend from the central nervous system exploring target organs and muscles and selected ones are stabilized by a nerve growth factor while others shrink back.
    \item The immune system of vertebrates can produce an extremely large variety of antibodies. A specific antibody can be activated by the appropriate antigen, being therefore stimulated to proliferate.
\end{itemize}

Similar to variation and selection in a population during evolution, such exploratory processes allow for variation and selection within an organism during development~\cite{gerhart2007theory}. That is, a variety of outcomes is generated of which the ones will be selected that work best depending on the surrounding environment and condition. When these conditions change, different outcomes can be selected. The property of exploratory behaviour of some systems, like the ones presented above, enhance the robustness of systems because otherwise harmful mutations can be accommodated by selection of a more suitable state among the diversity of states that is generated by these systems. It also enhances evolvability because it eliminates the need of coevolution for some systems and reduce finding evolutionary innovations to a smaller number of mutations to some regulatory elements. Such systems are probably selected and kept by evolution because of their high robustness, evolvability, and adaptability.

\begin{figure}
    \centering
    \includegraphics[width=\columnwidth]{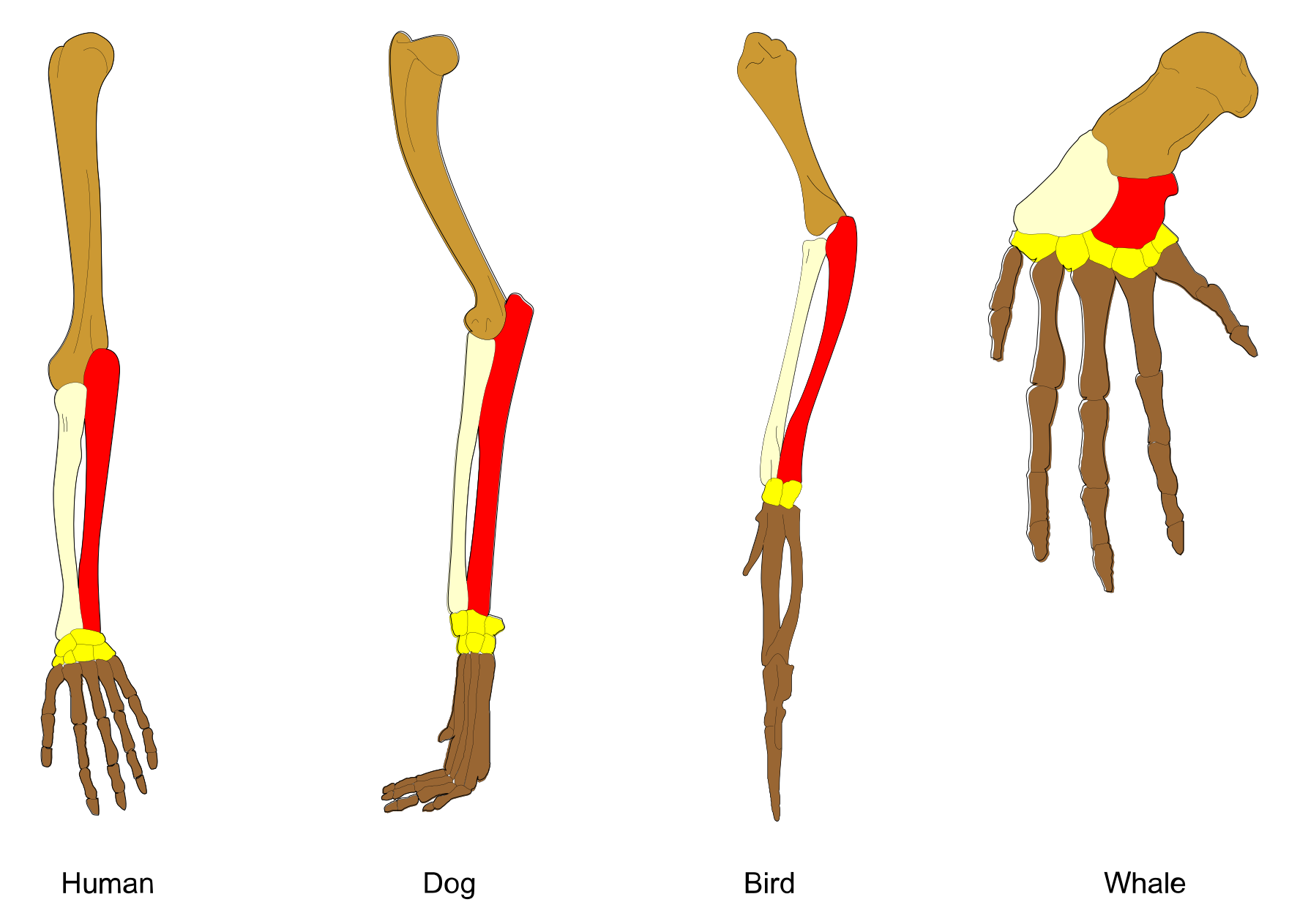}
    \caption{All mammals share the same basic bone structure in their forelimb.\protect\footnotemark}
    \label{fig:limbs}
\end{figure}

\footnotetext{By Volkov Vladislav Petrovich - Own work, CC BY-SA 4.0, \url{https://commons.wikimedia.org/w/index.php?curid=37704829}}

An elaborate example for the use of exploratory processes during development is the growth of a mammalian limb. While the skeleton has a modular structure and its growth is controlled by a modular developmental program that is highly conserved across all mammalian species, see Figure~\ref{fig:limbs}, the formation of the tissue---including blood vessels, muscles, and nerves---is accomplished by exploratory processes. This developmental mechanism, termed ``follow the bone'' principle in~\cite{kirschner2006plausibility}, demonstrates a sophisticated combination of exploitation (growth of skeleton) and exploration (growth of tissue), which provides flexibility on at least two scales. First, it helps to adjust body growth to the environmental conditions the organisms is exposed to, e.g., the food supply. Second, it facilitates the evolution of limbs because adaptions of the skeleton can be accomplished by a small number of mutations in some regulatory elements and tissue growth processes need not to coevolve.

\begin{figure}
    \centering
    \includegraphics[width=\columnwidth]{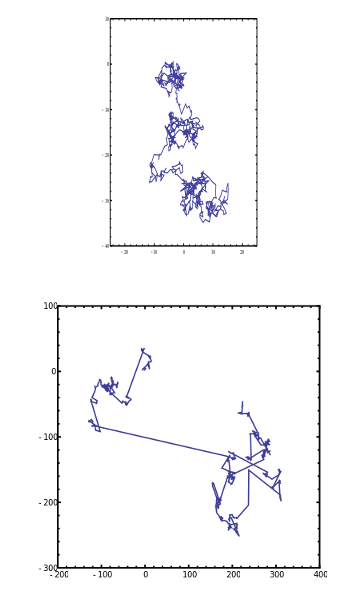}
    \caption{Brownian motion (top) vs. L\'evy flight (bottom). With the same number of steps, a L\'evy flight visits more different sites and returns less frequently to already visited sites than Brownian motion.\protect\footnotemark}
    \label{fig:levy}
\end{figure}

\footnotetext{By User:PAR - Own work, Public Domain, \url{https://commons.wikimedia.org/w/index.php?curid=9569860}}

Exploration also plays an important role when animals have to find randomly located objects, e.g., when searching for food in unknown territories or for mating partners for sexual reproduction. An interesting aspect of the search strategies is the distribution of travel distance per unit time. Model studies~\cite{viswanathan1999optimizing,bartumeus2002optimizing} have shown that heavy-tailed, scale-free distributions such as L\'evy flights~\cite{mandelbrot1982fractal}, where large jumps occur much more frequently than in Brownian motion, are advantageous under various natural environmental conditions. L\'evy flights minimize the probability of returning to the same site again and maximize the number of newly visited sites~\cite{viswanathan1999optimizing}, see Figure~\ref{fig:levy}. The applied search strategies affects the organism's flexibility as it determines how fast the organism can aquaint itself with unknown territory. For the importance of scale-free distributions in the context of system flexibility, see also Section~\ref{sec:weak}.

\subsection{Degeneracy and neutral spaces}
\label{sec:degeneracy}

Two or more systems, system configurations, or modules are called \emph{degenerate} if they have the same functions under certain conditions but distinct functions in other conditions~\cite{edelman2001degeneracy}. Degeneracy is different from \emph{redundancy} which means that the systems or modules have the same functions under all conditions, i.e., they are effectively copies of each other.

Given a task, a \emph{neutral space} is a collections of systems, system configurations or modules that perform the task equally well~\cite[p. 195]{wagner2007robustness}. With regard to the given task, the elements of the neutral space are equivalent. The set~$\Gamma(S,T)$ of all system configurations for which system $S$ can perform task $T$, which we have defined in Definition~\ref{defi:min_reco_cost}, is a neutral space. A \emph{neutral modification} refers to a transition from one element in a neutral space to another element in the same neutral space.

Degeneracy is present on many levels in biological systems~\cite{edelman2001degeneracy,mason2015hidden}.
Examples include different nucleotide sequences in the genetic code encoding the same polypeptide,
functionally equivalent alleles in genes, and different patterns of neural architecture being functionally equivalent. Degeneracy can also be observed in perceptual-motor systems~\cite{seifert2016neurobiological}. In inter-animal communication, there is a large number of ways by which the same message can be transmitted. Further, there are multiple acoustically diverse and temporally distributed cues that encode the same sound, and different tongue configurations that lead to the similar acoustic outputs~\cite{winter2014spoken}. For an extensive list of further examples, we refer the interested reader to~\cite{edelman2001degeneracy}.

For degeneracy to occur it requires a certain level of system complexity. Intuitively, the reason is that effective degeneracy needs sufficiently many structurally different parts in a system that can interact in sufficiently many different ways to yield the same output or functional result~\cite{edelman2001degeneracy}. Information-theoretical analyses of the relationship between degeneracy and complexity can be found in~\cite{tononi1994measure,tononi1996complexity,tononi1999measures}.

Degeneracy is pervasive in biological systems since natural evolution is likely to bring forth degenerate systems. There are several reasons for this. First of all, degeneracy is a source of robustness as it lowers the probability that a mutation is lethal~\cite{kirschner1998evolvability,draghi2010mutational,whitacre2010degeneracy}. Evolutionary search through mutation is initially undirected and in consequence will most likely find frequent solutions that can be realized in many different ways. Hence, problems with degenerate solutions tend to have a large neutral space. These natural neutral spaces contain fragile regions, where a search driven by mutation is likely to leave the space, and robust regions, where it is likely that the evolutionary search will remain in the neutral space. Through evolution by natural selection, some biological systems have come to reside in regions of high robustness~\cite[p.215f]{wagner2007robustness}. For such systems, most mutations will be neutral, i.e., they do not change a well-defined aspect of a biological system's function in a specific environment and genetic background~\cite[p. 224]{wagner2007robustness}.

\begin{figure}
    \centering
    \includegraphics[width=8.5cm]{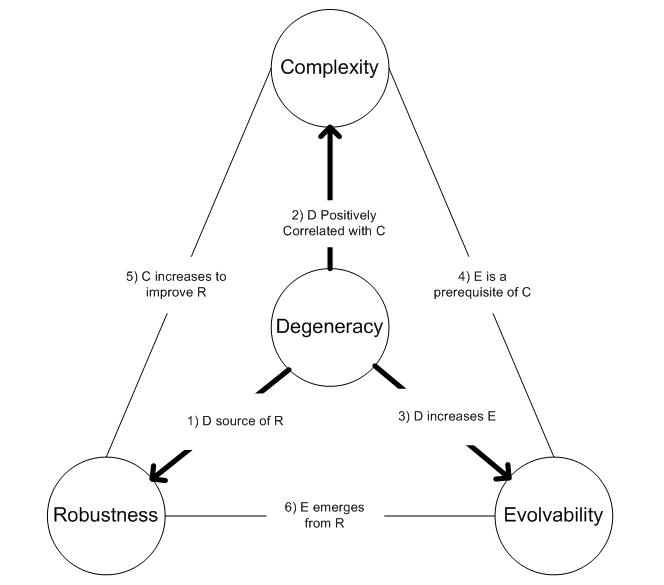}
    \caption{Relationship between degeneracy, complexity, robustness and evolvability according to~\cite{whitacre2010degeneracy}.\protect\footnotemark}
    \label{fig:degeneracy}
\end{figure}

While degeneracy is a source of robustness, it does also increases evolvability. Although neutral mutations do not change primary functions, they can change other system features and thus the phenotypic neighbourhood of genotypes. This allows organisms to evolve to regions of neutral spaces where most mutations will be neutral but few non-neutral mutations have a higher chance to trigger sudden useful phenotypic change. Another effect are exaptations~\cite{gould1982exaptation} which are organismal features that become adaptions only long after they arise.  Under certain conditions, it is shown that these effects confer systems more robustness as well as a greater evolvability, as more phenotypes are available than it would be possible only with non-neutral mutations~\cite{draghi2010mutational,whitacre2010degeneracy}. In consequence, degeneracy can be understood as both a prerequisite and a product of evolution and specifically natural selection~\cite{edelman2001degeneracy}. The interplay between degeneracy, complexity, evolvability, and robustness in biological systems is summarized in Figure~\ref{fig:degeneracy}.  

\footnotetext{By BioWikiEditor - Own work, CC BY-SA 3.0, \url{https://commons.wikimedia.org/w/index.php?curid=14948898}}

While degeneracy is able to increase evolvability, redundancy is not~\cite{whitacre2010degeneracy2}. Moreover, neither increasing neutral network exploration, or changing neutral network topology, is able to guarantee that evolvability will be increased. Only in degenerate systems evolvability can be increased by these means, and robustness or other properties alone do not guarantee better evolvability, as degenerate systems are structurally different and have different neutral networks and thus adaptability possibilities, whereas redundant systems, although they improve robustness, are the same, and will not offer a selective advantage when compared to each other in face of environmental changes or mutations~\cite{whitacre2010degeneracy2}.

\subsection{Weak links}
\label{sec:weak}

A link between two components of a system is called \emph{weak} when its addition or removal does not change the mean value of a target feature of the system in a statistically discernible way~\cite{berlow1999strong}. Note that `weak link' in this section has a different meaning than in Section~\ref{sec:wrl} on weak regulatory linkage.

Weak links are present in most biological systems on all scales of biological organization, from molecules and the cell, passing through systems in the organism such as immunological networks, muscle nets, and the neuro-glial network, reaching up to whole ecosystems~\cite{csermely2009weak}. Weak links also occur in social interaction networks and man-made systems such as the internet~\cite{csermely2009weak}.

In general, weak links come along with certain topological properties when considering systems and processes as networks. One of the recurring topological properties is scale-freeness, which means that the probability of a node having $k$ links decreases by a factor of $k^{-\alpha}$, with~$\alpha$ between $0$ and $3$~\cite{barabasi2004network}. More generally, there are many nodes with few connections and only a few nodes, called hubs, that are connected to a large number of other nodes. These hubs are generally connected to nodes with few links instead of to other hubs, which is called disassortative property. Another property present in many of these networks is the small-world effect, that describes networks where any two nodes are connected by a relatively short path. Motifs, modules, and hierarchy are also frequently observed, see also Section~\ref{sec:hierarchy} and Section~\ref{sec:modularity}. Two examples of biological systems exhibiting the network topological properties discussed above are protein-protein networks and metabolic networks.

In protein-protein networks, that model interactions between proteins, most proteins participate in only a few interactions, whereas a few proteins participate in dozens (scale-free topology)~\cite{barabasi2004network}. In S. Cerevisiae, only $~10$ percent of the proteins with less than 5 links are essential, but this fraction increases to over 60 percent for proteins with more than 15 interactions, which demonstrates the importance of hubs. Furthermore, highly interacting proteins have a smaller evolutionary distance to their orthologues. Hence, hubs are evolutionary conserved.

In metabolic networks, nodes are substrates/metabolites, and edges are enzyme-catalyzed biochemical reactions. Most metabolic substrates participate in only one or two reactions, but a few, such as pyruvate or coenzyme A, participate in dozens and function as metabolic hubs. Moreover, most reactions have quite small fluxes (the amount of substrate that is being converted to a product within a unit of time), coexisting with a few reactions with extremely high flux values.

In complex systems that are scale-free and disassortative most links are weak. As they do not connect hubs, the addition or removal of some of those links does not affect the system in a significant way. Also the strentgh of interactions (for example, the flux of a metabolic reaction in metabolic networks) contributes to the weakness of many links. Yet, in their totality, weak links play an important role for the adaptability and robustness of the system in several ways. While conserved hubs and modules are difficult to be modified, the weak links can easily change to adapt to new conditions or form different connection patterns to restore a certain system function. 

The importance of weak interactions between species in ecological communities has been demonstrated by taking the example of a simple rocky-intertidal food web system~\cite{berlow1999strong}. When there is only a predator and a prey, the interaction between them is always strong. However, when there is a second prey that interacts with the original predator and prey, the indirect interaction effect of the original predator and prey varies according to predator density and the density of the second prey, which means that in some situations it is strong and in other situations it is weak. Ultimately, when the direct effect between predator and prey is weak, then it is more susceptible to the indirect effect involving other components of the system, which also generates variation.

\begin{figure}
    \centering
    \includegraphics[width=8cm]{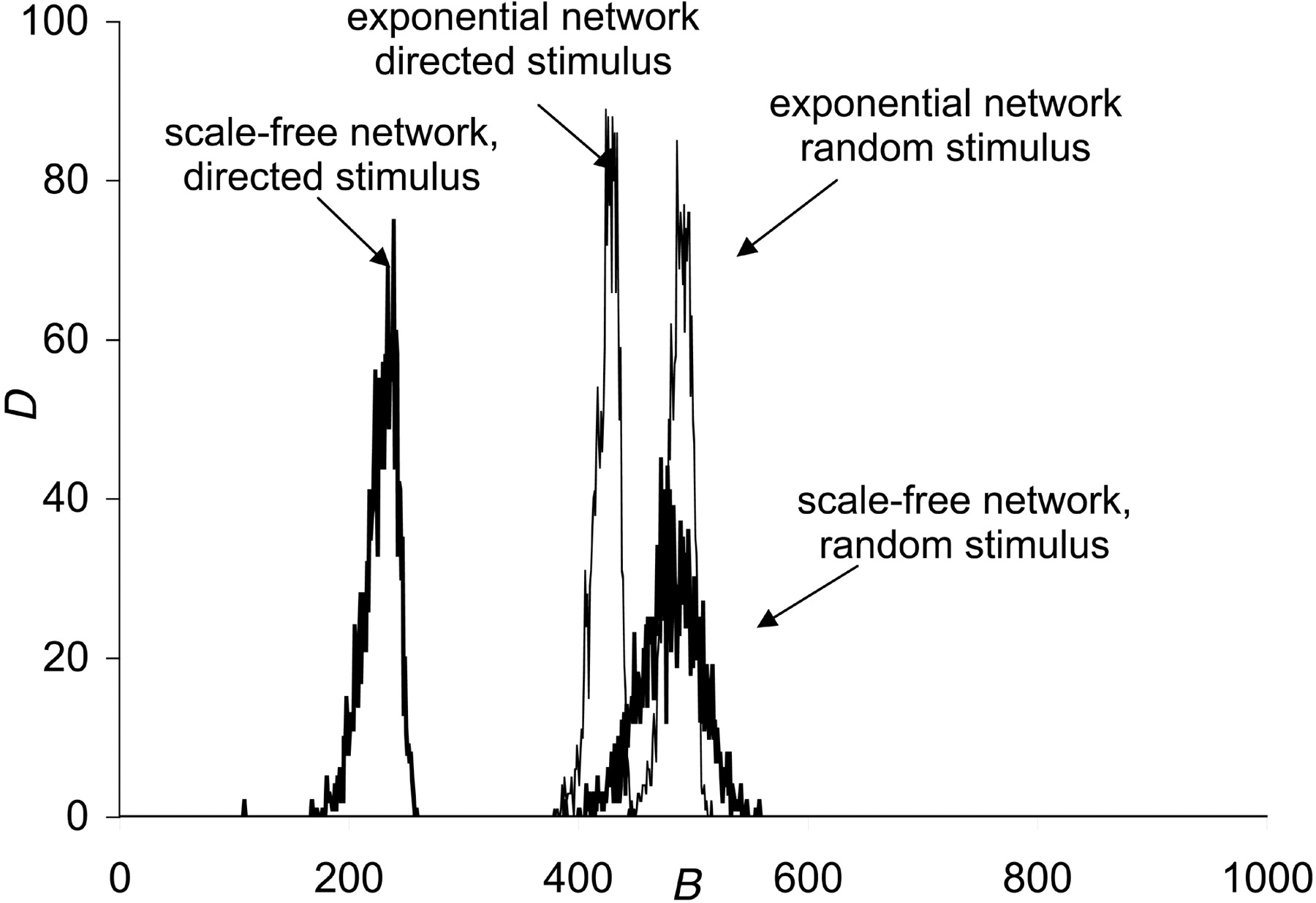}
    \caption{Histograms of the number of changed nodes B needed to change the state of a network with two randomly selected functional states in a network of 1,000 nodes. For details, see~\cite{baryam2004response}. Copyright (2004) National Academy of Sciences.}
    \label{fig:weak_links}
\end{figure}

The effect of weak links on system adaptability has been demonstrated in~\cite{baryam2004response}. The paper studies complex networks that have a high sensitivity to determined kinds of stimuli while still retaining robustness for more general/random stimuli. Complex systems with these two properties are more robust to environmental changes and at the same time more adaptable. Results of simulations show that both random and scale-free network topologies have a similar behavior when random stimuli are used, but scale-free networks require much less nodes to be affected before changing to another state when stimuli directed to the nodes with higher connectivity are used. In addition, if a value for $\alpha$ (the scale parameter) similar to the ones found in networks in nature is used, then it is possible to build networks that require four or even only one node with the highest connectivity to be affected for the entire network to change its state, whereas much more nodes need to be affected by random stimuli for that to happen, see Figure~\ref{fig:weak_links}. These networks, thus, have a high sensitivity to the right stimuli while still retaining robustness to random stimuli.

\section{Summary and outlook}
\label{sec:outlook}

Based on the abstract notion of task-performing system, we have presented a general formalism to describe flexibility problems from various disciplines, e.g., evolutionary biology, machine learning, or computer-integrated manufacturing. Moreover, we have discussed the elements of flexibility, which are universal design features of living systems that facilitate flexibility. Based on the elements of flexibility, there are a number of possible research directions.

\paragraph{Incorporating the elements.} For applied flexibility problems, a bio-inspired research agenda would deal with the question in which concrete form the elements of flexibility could be integrated in a given task-performing system to improve its flexibility in a given task context. General guiding questions are: 
\begin{enumerate}[label=\Roman*.]
    \item Which of the elements of flexibility are already present in the system under consideration?
    \item Are the elements of flexibility which are present in the system already used to increase the system's flexibility?
    \item Could further elements of flexibility be incorporated into the system?
    \item Is the interplay of the elements of flexibility understood for the system under consideration?
\end{enumerate}

Studying the elements of flexibility in this form for cyber-physical production systems can contribute to the emerging research frontier \emph{biologicalization in manufacturing}, which refers to ``the use and integration of biological and bio-inspired principles, materials, functions, structures and resources for intelligent and sustainable manufacturing technologies and systems with the aim of achieving their full potential''~\cite{byrne2018biologicalisation}.

\paragraph{Quantifying the contribution of an element.} For a comprehensive understanding of the elements of flexibility, a particularly important research aspect is to quantify the contribution of an element to the flexibility of a given type of task-performing system in a given task-context. This requires in the first place a measure to quantify to which degree an element of flexibility is present or not. For hierarchy, modularity, degeneracy, and weak links, such measures are available in the literature, see Section~\ref{sec:elements} and the references there. With regard to exploratory behavior, investigating the contribution of exploratory behavior is closely related to the exploitation-exploration-dilemma~\cite{berger2014exploration,ghavamzadeh2015bayesian}. To be able to say to which degree weak regulatory linkage is present or not, it first requires a more rigorous formal definition that is not available in the literature yet.

Once a measure is available that quantifies the presence of an element, one can conduct numerical experiments in a given task context to measure the effect the presence of the element has on the flexibility measure. To our knowledge, there are so far only a few works that follow such an approach~\cite{clune2013origins,mengistu2016evolutionary,sotto2022pole}.

\paragraph{Benchmark flexibility problems.}
Clearly, a direct extensive study of the elements of flexibility for systems that have to perform physical tasks is often impractical due to high experimental cost. This requires to resort to suitable computer experiments based on numerical simulations where only final outcomes are validated in physical experiments. But also for realistic computation and learning tasks, a direct systematic flexibility optimization can be too costly. We thus believe that research on system flexibility optimization and the elements of flexibility can benefit tremendously from properly devised \emph{benchmark flexibility problems}. An important criterion for the tasks in such a benchmark flexibility problem is that solutions are known an each single task is cheap to solve in terms of required training data or numerical simulations. Moreover, the task context should consists of well-established benchmark tasks. This allows to focus complexity on flexibility aspects. Providing domain-specific benchmark flexibility problems is already a valuable research contribution.

%\begin{conjecture}
%The capability of machine learning models to learn and build compositional structures is key to achieving human-like artifical intelligence~\cite{lake2017building}.
%\end{conjecture}

%\begin{conjecture}
%The study of embryogenesis can inspire and inform us on how to create more automated development processes for cyber-physical production systems.
%\end{conjecture}

%\begin{conjecture}
%The more complex technical systems become and the more machine learning, in particular deep learning, is part of the systems and involved in the development process, the %more the design of the systems can become a black-box. Methods developed to identify modules in biological systems can help to identify modules in such technical system %to open up the black-box again.
%\end{conjecture}

%\begin{conjecture}
%Algorithms that have been developed to understand the evolutionary origins of modularity in biological systems can be used the evolve modularly organized technical %systems.
%\end{conjecture}

%\begin{conjecture}
%The study of weak regulatory linkage can help do develop more complex and flexible networks of interacting intelligent agents.
%\end{conjecture}

\paragraph{Achknowledgement}
We would like to thank Martin Kretschmer, Christoph Quix (both Fraunhofer FIT) and Christopher Vahl (Fraunhofer SCAI) for valuable feedback. Part of this work was carried out during the tenure of an ERCIM ‘Alain Bensoussan’ Fellowship Programme.

\bibliographystyle{IEEEtran}
\bibliography{newcybernetics}

\appendix

\section{Evolvability of logic circuits}
\label{sec:boolean_circuit_evolvability}

One of the striking features of evolution is the appearance of novel structures in organisms. The origin of the ability to generate novelty is one of the main mysteries in evolutionary theory. The molecular mechanisms that enhance the evolution of novelty were integrated by Kirschner and Gerhart in their theory of facilitated variation (FV)~\cite{gerhart2007theory}. This theory suggests that organisms have a design that makes it more likely that random genetic changes will result in organisms with novel shapes that can survive. The paper~\cite{parter2008faciliated} demonstrates by means of computer simulations of two well-studied model systems, logic circuits and RNA secondary structure, that FV can spontaneously emerge. We recapitulate their main results on logic circuits using the formalism developed in Section~\ref{sec:flexibility} and discuss which elements of flexibility play a role here.

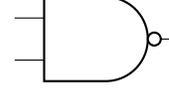
\begin{figure}
    \centering
    \begin{circuitikz}
                \draw
                (0,0) node {} % blank
                (1.5, 1) node[ieeestd nand port] (mynand) {};
    \end{circuitikz}
    \caption{IEEE shape symbol of NAND gate.}%
    \label{fig:nand_gate}
\end{figure}

Before we get to the formalism, let us first describe the essence of the flexibility problem considered in~\cite{parter2008faciliated}. A \emph{Boolean expression} is a formula that evaluates to~$1$~(true) or~$0$~(false). An example for a Boolean expression with four variables~$x_1,x_2,x_3,x_4 \in \{0,1\}$ is
\begin{align}
     t(x_1,x_2,x_3,x_4) = (x_1\; \mathrm{ AND }\; x_2)\; \mathrm{ OR }\; (x_3\; \mathrm{ EQ }\; x_4).
\end{align}
A Boolean expression including Boolean variables is also called \emph{Boolean function}. 
It is known that every Boolean function can be equivalently constructed from a combination of NAND gates.
The NAND gate is defined by the truth table given in Table~\ref{tab:nand_gate} and its standard shape symbol is depicted in Figure~\ref{fig:nand_gate}.
\begin{figure}
    \begin{tabular}{|c|c|c|}
                \hline
                $x_1$ & $x_2$ & $\mathrm{NAND}(x_1,x_2)$\\
                \hline
                $0$ & $0$ & $1$\\
                $0$ & $1$ & $1$\\
                $1$ & $0$ & $1$\\
                $1$ & $1$ & $0$\\
                \hline
    \end{tabular}
    \caption{Truth table of NAND gate.}%
    \label{tab:nand_gate}
\end{figure}

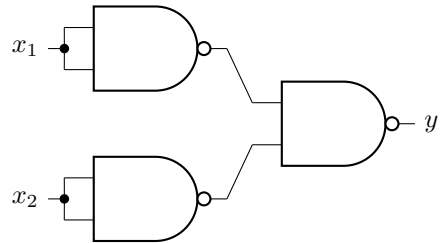
\begin{figure}
    \centering
    \begin{circuitikz}
        \node[ocirc,fill=black](split1) at (0.42,2) {};
        \node[ocirc,fill=black](split2) at (0.42,0) {};
        \draw
        (-0.1,2) node (in1) {$x_1$}
        (-0.1,0) node (in2) {$x_2$}
        (1.5,2) node[ieeestd nand port] (mynand1) {}
        (1.5,0) node[ieeestd nand port] (mynand2) {}
        (4,1) node[ieeestd nand port] (mynand3) {}
        (5.3,1) node (out1) {$y$}
        (in1.east) -- (split1.west)
        (in2.east) -- (split2.west)
        (split1.north) -- (mynand1.in 1)
        (split1.south) -- (mynand1.in 2)
        (split2.north) -- (mynand2.in 1)
        (split2.south) -- (mynand2.in 2)
        (mynand1.out) -- (mynand3.in 1)
        (mynand2.out) -- (mynand3.in 2);
    \end{circuitikz}
    \caption{NAND construction of the OR gate.}
    \label{fig:OR_construction}
\end{figure}

\noindent As an example, consider the OR gate. It can be written as
\begin{align}
     x_1\; \mathrm{ OR }\; x_2 =  (x_1\; \mathrm{ NAND }\;x_1) \; \mathrm{ NAND }\; (x_2\; \mathrm{ NAND }\;x_2). 
\end{align}
The corresponding circuit is depicted in Figure~\ref{fig:OR_construction}.

\noindent The flexibility problem that is addressed in~\cite{parter2008faciliated} is essentially the following.

\begin{definition}[Logic circuit flexibility problem (LCFP)]
\label{defi:flexibility_problem}
Given Boolean functions~$t_1, t_2,\dots,t_n$, find a template Boolean function build solely from NAND gates such that its structure allows for the following:
\begin{enumerate}
    \item For any Boolean function $t_i$, it takes only few modifications to the template to have a construction of $t_i$ based on NAND gates (low reconfiguration cost).
    \item It is easy to find the necessary modifications (low adaption cost).
\end{enumerate}
Moreover, we wish the reconfiguration and adaption cost to be low for as many different Boolean functions as possible (high task diversity).
\end{definition}

\noindent Definition~\ref{defi:flexibility_problem} captures the essence of the flexibility problem. A number of details are still missing to have a problem formulation that is completely precise and detailed enough that a computer can solve it. We use the flexibility formalism to guide us through the procedure of filling the gaps. In the end, we will have a precise picture of the exact version of LCFP considered in~\cite{parter2008faciliated} and the system used to solve it.  

\subsection{Formal description of the tasks}
\label{sec:task_description}
Reproducing the truth values of a Boolean expression in $d$ Boolean variables~$x=(x_1,\dots,x_d)$  can be described as a task
\begin{align}
     T = (E,F,t,P_T)
\end{align}
with the following elements. The environment is given by the space of all possible values (or states) a vector of $d$ Boolean variables can take,~$E=\{0,1\}^d$. All states are feasible, so $F=E$, and all states are possible initial states, hence $X=E$. The goal is now to map an input state $x \in X$ to a target state in
\begin{align}
     Y = \{ (x_1,1,1,1): x_1 \in \{0,1\}  \}.
\end{align}
Note that $Y$ is one of $2^{d-1}$ subspaces in $E$ that are equivalent to $\{0,1\}$. The particular choice of $Y$ is arbitrary and we may simply identify $Y=\{0,1\}$. The goal is formally given by a Boolean function
\begin{align}
     t: \{0,1\}^d \to \{0,1\}.
\end{align}
The performance (or fitness) of a given Boolean function $u:  \{0,1\}^d \to \{0,1\}$ in reaching the goal is measured by the fraction of outputs where $u$ coincides with~$t$,
\begin{align}
     \tilde P (u,t) = \frac{|\{x \in E: u(x) = t(x) \}|}{|E|}.
\end{align}

To be able to read adaptability and reconfigurability numbers from~\cite{parter2008faciliated}, we consider a task to be performed successfully by a Boolean function $u$ if we have~$\tilde P(u,t) \geq \varepsilon$, where the specific choice of $\varepsilon \in [0,1]$ is discussed in Section~\ref{sec:further_details}. Hence, the ultimate performance measure  is given by
\begin{align}
    P_T(u) = \begin{cases}
     1, \text{ if } \tilde P(u,t) \geq \varepsilon,\\
     0, \text{ otherwise}.
    \end{cases}
\end{align}
We discuss the specific task context considered in the flexibility problem in Section~\ref{sec:flexibility_problem}.

\subsection{Formal description of the system}

In~\cite{parter2008faciliated} a standard genetic algorithm (GA) is used to evolve a population of logic circuits consisting of NAND gates towards varying Boolean target functions~$t$. The population size is $N_{\mathrm{pop}} = 5000$. We consider both together---the logic circuit and the GA---as a self-adapting system $S$ where the GA operates as the learning system. To be more precise, the system $S$ consists of the following elements:
\begin{enumerate}
    \item A logic circuit consisting of
    \begin{enumerate}
        \item $d$ input gates ($d=4$),
        \item $M$ NAND gates  ($M=12$),
        \item one output gate.
    \end{enumerate}
    The input gates can be connected to more than one non-input gate. A NAND gate can be connected to more than one of the other NAND gates and to the output gate.
    This is the \emph{phenotype} in the language of evolutionary computation.
    \item A binary encoding $\gamma$ of the circuit wiring. This is the \emph{genotype} in the language of evolutionary computation and what we consider the \emph{system configuration} to be adapted.
    \item A genotype-phenotype-mapping that translates the genotype into an actual wiring of the logic circuit.
    \item An array of length $N_{\mathrm{pop}}$ that contains alternative system configurations.
    \item The genetic algorithm. 
\end{enumerate}

\noindent The genetic algorithm operates on the array of alternative genotypes. Every generation it evaluates the performance of all genotypes/system configurations in the array and selects the best-performing as the current configuration $\gamma$ of the system. The system then realizes an algorithm $S(\gamma)$ that compute the $d$-variate Boolean function specified by the wiring of the NAND gates.

It remains to describe the structure of the binary encoding and the resulting system configuration space $\Gamma(S)$. The encoding consists of $M$ ``gate genes'' and one ``output gene''. A gate gene has a length of
\begin{align}
    2m = 2\lceil\log_2(M+d)\rceil=8
\end{align}
bits where the first $m$ bits encode which input or NAND gate is connected to the first input of the gate, and the second $m$ bits encode the wiring of the second input. The output gate has a length of $m=4$ bits. Hence, the binary encoding has a length of $B=4(2M+1)$ bits and the system's configuration space is given by
\begin{align}
    \Gamma(S) = \{0,1\}^B.
\end{align}

\subsection{Adaption and reconfiguration cost}

Given a current system configuration $\gamma \in \Gamma(S)$ and a new task $T$, the adaption cost are the number of generations it takes the genetic algorithm to find a configuration $\gamma' \in \Gamma(S)$ such that $P_T(S(\gamma')) = 1$. The reconfiguration cost are the number of mutations that have to be applied to get from the previous configuration $\gamma$ to the new configuration $\gamma'$ (i.e., the Hamming distance between~$\gamma$ and~$\gamma'$).
 
\subsection{The flexibility optimization problem in detail}
\label{sec:flexibility_problem}

Let us now describe one variant of LCFP that is actually addressed in \cite{parter2008faciliated} using the language of the flexibility formalism. The considered task context consists of tasks $T$ that all share the same environment $E$ but have different goals~$t: X \to Y$ of the form
\begin{align}
   t(x_1,x_2,x_3,x_4) = f(g(x_1,x_2), h(x_3,x_4))    
\end{align}
where $g,h \in \{\mathrm{EQ}, \mathrm{XOR}\}$ and $f \in \{\mathrm{AND},\mathrm{OR}\}$. More specifically, we have a training task context
\begin{align}
    W^{\mathrm{train}} = ((T_1, T_2, T_3), Q)
\end{align}
and a test task context $W^{\mathrm{test}}$ consisting of twenty equally probable tasks where the goals are such that either $f$, $g$, or $h$ is a Boolean function not present in~$T_1, T_2, T_3$. What is now examined by means of a computer experiment in the paper, see \cite[Fig. 6D]{parter2008faciliated}, is the effect of two different task histories generated from the training context~$W^{\mathrm{train}}$ on the adaptability and reconfigurability of the system~$S$ in the test context~$W^{\mathrm{test}}$. Hence, LCFP is solved for the Boolean expressions given in $W^{\mathrm{test}}$. We discuss their results in the next section

\subsection{Adaptability and reconfigurability results}

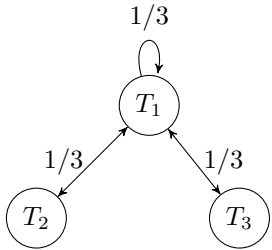
\begin{figure}[b]
    \centering
    \begin{tikzpicture}[->]
        \node[circle,draw] (T_1) at (10,10) {$T_1$};
        \node[circle,draw] (T_2) at (8.5,8.5) {$T_2$};
        \node[circle,draw] (T_3) at (11.2,8.5) {$T_3$};

        \path (T_1) edge[<->] node[above,left] {$1/3$} (T_2);
        \path (T_1) edge[<->] node[above,right] {$1/3$} (T_3);
        \path (T_1) edge[loop above] node {$1/3$} (C);
  \end{tikzpicture}
    \caption{Transition probabilities between tasks in the MVG scenario.}
    \label{fig:mvg_graph}
\end{figure}

In the first case, called \emph{fixed goal (FG) scenario} in the paper, the system is only presented one task $T_1$ for a certain number of generations $L$ (formally, the probability measure $Q$ is chosen such that $T_1$ has probability $1$). In the second case, called \emph{modularly varying goals (MVG) scenario}, the system is presented a random task history that results from performing a random walk between three tasks $T_1, T_2, T_3$ according to the transition probabilities in Figure~\ref{fig:mvg_graph}, where the task is switched every $E=20$ generations. This leads to two task histories $H_n^{\mathrm{FG}}$ and $H_n^{\mathrm{MVG}}$ of length $n=L/E$.

Let $T$ be a random task from $W^{\mathrm{test}}$. Given our choice of the performance threshold~$\varepsilon$, see Section~\ref{sec:further_details}, the numerical results in \cite[Fig. 6D]{parter2008faciliated} show that
\begin{align}
 \mathbb{E}[c^{\textrm{ada}}(S,H_n^{\mathrm{FG}}, T)] > 10,
\end{align}
while
\begin{align}
 \mathbb{E}[c^{\textrm{ada}}(S,H_n^{\mathrm{MVG}}, T)] = 3.
\end{align}
Moreover, the reconfiguration cost in the MVG scenario are on average five times smaller than in the FG scenario, see~\cite[last paragraph on p.3]{parter2008faciliated}.

Though we do not change the design of the system $S$ or the design of the configuration space $\Gamma(S)$, one may still consider the training phase in the MVG scenario as a system optimization that improves the array of alternative system configurations and the initial configuration in the testing phase by locating them in a beneficial region of the configuration space $\Gamma(S)$. In that sense, the FG and the MVG scenario lead to two different systems $S^{\mathrm{FG}}$ and $S^{\mathrm{MVG}}$, where the latter is at least three times more adaptable,
\begin{align}
 \alpha_n(S^{\mathrm{MVG}}, W^{\mathrm{test}}) > 3\, \alpha_n(S^{\mathrm{FG}}, W^{\mathrm{test}}),
\end{align}
and five times easier to reconfigure on average,
\begin{align}
 \rho^{\mathrm{avg}}(S^{\mathrm{MVG}}, W^{\mathrm{test}}) > 5\, \rho^{\mathrm{avg}}(S^{\mathrm{FG}}, W^{\mathrm{test}}).
\end{align}

\subsection{Elements of flexibility}

The reason for the better adaptability and reconfigurability in the MVG scenario is that the switching between the tasks in the history $H_n^{\mathrm{MVG}}$ lets the genetic algorithm search in a region of the configuration space $\Gamma(S)$ where the corresponding wirings between the NAND gates reflect the hierarchical and modular structure of goals, which is not the case in the FG scenario. To see how also weak regulatory linkage appears in the MVG scenario, we refer to the explanations in~\cite{parter2008faciliated}.

\subsection{Further details}
\label{sec:further_details}
\paragraph{Choice of threshold $\varepsilon$.}
To exclude that the same results could be reached by random guessing, the authors consider a normalized fitness measure given by
\begin{align}
 F_N = \frac{F - F_r}{1- F_r},
\end{align}
where $F = \max_{\gamma_1,\dots,\gamma_{N_{\mathrm pop}}} P_T(S(\gamma_i))$ is the maximal fitness in the evolved population and $F_r$ the average maximal fitness in a population of random genotypes of the same size $N_{\mathrm{pop}}$. The results in \cite[Fig. 6D]{parter2008faciliated} are specified for the normalized fitness. Since $F_r$ is independent of the considered system $S$, we may consider $F_r$ as a constant such that $F_N$ is proportional to $F$. We choose $\varepsilon$ such that
\begin{align}
 F = F_r + (1-F_r) F_N > \varepsilon
\end{align}
if $F_N > 0.8$.

\end{document}